\documentclass[12pt]{article}

\usepackage{arxiv}

\title{Breaking Feedback Loops in Recommender Systems with Causal Inference}

\date{}

\author{Karl Krauth\\
       University of California Berkeley\\
       \texttt{karlk@berkeley.edu}
       \And
       Yixin Wang\\
       University of Michigan\\
       \texttt{yixinw@umich.edu}\\
       \And
       Michael I.\ Jordan\\
       University of California Berkeley\\
       \texttt{jordan@cs.berkeley.edu}}

\renewcommand{\headeright}{}
\renewcommand{\undertitle}{}
\renewcommand{\shorttitle}{Breaking Feedback Loops in Recommender Systems with Causal Inference}

\usepackage[colorlinks,linktoc=all]{hyperref}
\hypersetup{linkcolor=black}
\hypersetup{urlcolor=violet}
\usepackage{amsthm}

\input{preamble/preamble}

\newacronym{ELBO}{elbo}{evidence lower bound}
\newacronym[shortplural=vae]{VAE}{vae}{variational autoencoder}
\newacronym{GMVAE}{gmvae}{Gaussian mixture VAE}
\newacronym{GMM}{gmm}{Gaussian mixture model}
\newacronym{AU}{au}{active units}
\newacronym{PPCA}{ppca}{probabilistic principal component analysis}
\newacronym{KL}{kl}{Kullback-Leibler}
\newacronym{LDA}{lda}{latent Dirichlet allocation}
\newacronym{SVI}{svi}{stochastic variational inference}
\newacronym{ICNN}{icnn}{input convex neural network}

\newacronym{ML}{ml}{maximum likelihood}
\newacronym{MLE}{mle}{maximum likelihood estimate}
\newacronym{MCMC}{mcmc}{Markov chain Monte Carlo}
\newacronym{HMC}{hmc}{Hamiltonian Monte Carlo}

\newacronym{LBFGS}{l-bfgs}{limited-memory Broyden-Fletcher-Goldfarb-Shanno}
\newacronym{ADVI}{advi}{automatic differentiation variational inference}
\newacronym{NUTS}{nuts}{No-U-Turn sampler}

\newacronym{GLM}{glm}{generalized linear model}

\newacronym{IF}{if}{influence function}

\newacronym{PF}{pf}{Poisson factorization}

\newacronym{MI}{mi}{mutual information}

\newacronym[\glsshortpluralkey={rpm}]
{RPM}{rpm}{reweighted probabilistic model}

\newacronym{NDCG}{ndcg}{normalized discounted cumulative gain}

\newacronym{MAP}{map}{mean average precision}

\newacronym{IDVAE-MT}{idvae-mt}{identifiable VAE via monotone
transport maps}

\newacronym{IDGMVAE-MT}{idgmvae-mt}{identifiable Gaussian mixture VAE via monotone
transport maps}

\newacronym{IDSVAE-MT}{idsvae-mt}{identifiable sequential VAE via
monotone transport maps}

\newacronym{IDVAE}{idvae}{identifiable VAE}

\newacronym{IDGMVAE}{idgmvae}{identifiable Gaussian mixture VAE}

\newacronym{IDSVAE}{idsvae}{identifiable sequential VAE}
\newacronym{IDMVAE}{idmvae}{identifiable mixture VAE}

\newacronym{CAFL}{cafl}{causal adjustment for feedback loops}

\newacronym{PROB-MF}{prob-mf}{probabilistic matrix factorization}

\newacronym{IPW}{ipw}{inverse probability weighting}
\newacronym{MNAR}{mnar}{missing-not-at-random}

\usepackage{tikz}
\usetikzlibrary{bayesnet}
\usepackage{pgfplots}
\pgfplotsset{compat=newest}
\pgfplotsset{plot coordinates/math parser=false}
\usepgfplotslibrary{statistics}
\usetikzlibrary{calc}
\pgfdeclarelayer{edgelayer}
\pgfdeclarelayer{nodelayer}
\pgfsetlayers{edgelayer,nodelayer,main}

\definecolor{hexcolor0xbfbfbf}{rgb}{0.749,0.749,0.749}

\tikzset{>=latex}
\tikzstyle{none}   = [inner sep=0pt]
\tikzstyle{line}   = [ thick, -, shorten <=1pt, shorten >=1pt ]
\tikzstyle{arrow}  = [ thick,  ->, shorten <=1pt, shorten >=1pt ]
\tikzstyle{ardash} = [ thick dotted, ->, shorten <=1pt, shorten >=1pt ]

\tikzstyle{empty}=[circle,opacity=0.0,text opacity=1.0,minimum width=4pt,minimum height=4pt]
\tikzstyle{box}=[rectangle,fill=white,draw=black]
\tikzstyle{filled}=[circle,fill=hexcolor0xbfbfbf,draw=black]
\tikzstyle{hollow}=[circle,fill=white,draw=black]
\tikzstyle{param}=[rectangle,fill=Black,draw=Black,inner sep=0pt,minimum width=4pt,minimum height=4pt]
\tikzstyle{paramhollow}=[rectangle,fill=White,draw=Black,inner sep=0pt,minimum
width=4pt,minimum height=4pt]

\DeclareRobustCommand{\mb}[1]{\ensuremath{\mathbf{\boldsymbol{#1}}}}

\DeclareMathOperator*{\argmax}{arg\,max}
\DeclareMathOperator*{\argmin}{arg\,min}

\crefname{lemma}{lemma}{lemmas}
\Crefname{lemma}{Lemma}{Lemmas}
\crefname{thm}{theorem}{theorems}
\Crefname{thm}{Theorem}{Theorems}
\crefname{prop}{proposition}{propositions}
\Crefname{prop}{Proposition}{Propositions}
\crefname{assumption}{assumption}{assumptions}
\Crefname{assumption}{Assumption}{Assumptions}
\crefname{corollary}{Corollary}{Corollaries}
\Crefname{corollary}{Corollary}{Corollaries}
\crefname{exmp}{example}{examples}
\Crefname{exmp}{Example}{Examples}
\crefname{defn}{definition}{definitions}
\Crefname{defn}{Definition}{Definitions}
\creflabelformat{equation}{#1#2#3}
\crefname{equation}{eq.}{eqs.}
\Crefname{equation}{Eq.}{Eqs.}
\crefname{figure}{fig.}{figs.}
\Crefname{figure}{Fig.}{Figs.}

\newtheorem{thm}{Theorem} %
\newtheorem{defn}{Definition} %
\newtheorem{prop}[thm]{Proposition}
\newtheorem{assumption}{Assumption}

\newcommand{\mba}{\mb{a}}
\newcommand{\mbA}{\mb{A}}
\newcommand{\mbc}{\mb{c}}

\newcommand{\mbR}{\mb{R}}

\newcommand\dif{\mathop{}\!\mathrm{d}}

\newcommand{\E}[2]{\mathbb{E}_{#1}\left[#2\right]}

\newcommand{\cN}{\mathcal{N}}

\newcommand{\g}{\, | \,}
\newcommand{\s}{\, ; \,}

\newtheorem{corollary}{Corollary}

\usepackage[utf8]{inputenc} %
\usepackage{url}            %
\usepackage{amsfonts}       %
\usepackage{microtype}      %
\usepackage{doi}

\newtheorem{lemma}{Lemma}
\hypersetup{
pdftitle={Breaking Feedback Loops in Recommender Systems with Causal Inference},
pdfsubject={cs.IR, cs.LG, cs.CY, stat.ML},
pdfauthor={Karl Krauth, Yixin Wang, Michael I.~Jordan},
pdfkeywords={Causal Inference, Recommender System, Feedback Loops, Recommendation System, Homogenization},
}

\usepackage{amssymb}

\usepackage{bbm}
\usepackage{times}
\usepackage{adjustbox}
\usepackage{tcolorbox}
\usepackage{enumitem}
\usepackage[authoryear]{natbib}
\usepackage{booktabs,arydshln}
\usepackage{cuted}
\usepackage{wrapfig}
\usepackage{subcaption}
\usepackage{tcolorbox}

\usepackage[dvipsnames]{xcolor}

\definecolor{salmon}{RGB}{234,153,153}
\definecolor{cornflowerblue}{RGB}{100,149,237}
\hypersetup{
    colorlinks,
    linkcolor={black},
    citecolor={cornflowerblue},
    urlcolor={salmon}
}

\usepackage{libertine}

\begin{document}

\maketitle

\begin{abstract}
Recommender systems play a key role in shaping modern web
ecosystems. These systems alternate between (1) making recommendations
(2) collecting user responses to these recommendations, and (3)
retraining the recommendation algorithm based on this feedback.
During this process, the recommender influences the user behavioral
data that is subsequently used to update the recommender itself, thus creating a feedback loop.
Recent work has shown that feedback loops may compromise recommendation quality
and homogenize user behavior, raising ethical and performance concerns around deploying recommender
systems. To address these concerns, we propose the \textit{\gls{CAFL}}, an
algorithm that uses causal inference to break feedback loops for any loss-minimizing recommendation algorithms. The key observation is that a recommender
system does not suffer from feedback loops if it reasons about causal
quantities, namely the intervention distributions of recommendations
on user ratings. Moreover, we can calculate these intervention
distributions from observational data by adjusting for the
recommender system's predictions of user preferences. Using
simulated environments, we demonstrate that \gls{CAFL}
improves recommendation quality when compared to prior correction methods.
\end{abstract}

\section{Introduction}
\label{sec:introduction}
Recommender systems are deployed in dynamic environments. At each
time step, a recommender system makes recommendations and collects
user feedback. At the next time step, the recommendation algorithm is
updated based on this feedback. The process alternates
between these two steps, inducing a feedback loop: the recommendation
algorithm influences what user behavior data it observes; this data in
turn affects the recommendation algorithm, since the algorithm is trained on this
data. Over time, the issue is exacerbated: the recommender system is
trained on a growing set of data points that have been biased by
recommendations.

Uncontrolled feedback loops create negative externalities that are
shouldered by consumers and producers. For example, they compromise
recommendation quality as they bias the behavioral data collected by
the system. They also exacerbate homogenization effects
\citep{chaney2018algorithmic}: if a user interacts with an item early
on, the recommendation system is more likely to recommend similar
items at the expense of dissimilar items that the user might prefer. A
related issue is the ``rich-get-richer'' problem where items that are
popular early on are undeservedly recommended over newer items since
the recommender system has observed more data about them
\citep{chakrabarti2006influence, salganik2006experimental}.

\glsreset{CAFL} A naive way to break feedback loops is to
recommend random items. Such a system does not suffer from negative
feedback effects because its recommendations no longer depend on past
data, but it does not learn from user behavior and would
unacceptably degrade recommendation quality. So how can we
break feedback loops without making recommendations useless?

In this work, we study the \emph{causal} mechanism underlying the
recommendation process and propose the \textit{\gls{CAFL}}, an algorithm
that can provably break feedback loops in recommender systems.
Studying feedback loops with a causal lens leads to a key observation:
recommendation algorithms do not suffer from feedback loops if they
reason about causal quantities, namely the intervention distributions
of recommendations on user ratings. The reason is that a \textit{do}
intervention on a causal graph, by definition, breaks the connection
between the variable being intervened on (e.g., the recommendation)
and its normal causes (e.g., user
feedback)~\citep{pearl2009causality}.

\begin{figure}
\begin{subfigure}{.2\textwidth}
\vspace{48pt}
\centering
\begin{tikzpicture}[every node/.style={circle, minimum size=1.0cm}]
    \node[hollow] (a1) at (0,0) {$\mbA$};
    \node[hollow] (r1) [above =of a1] {$\mbR$};

    \path[arrow] (a1) edge[bend right=30] (r1);
    \path[arrow] (r1) edge[bend right=30] (a1);
\end{tikzpicture}
\caption{\label{fig:feedback}}
\end{subfigure}
\begin{subfigure}{.7\textwidth}
\begin{tikzpicture}[every node/.style={circle, minimum size=1.0cm}]
    \node[hollow] (a1) at (0,0) {$\mbA_1$};
    \node[hollow] (r1) [above =of a1] {$\mbR_1$};
    \node[hollow] (t1) [right =of a1] {$\hat{\Theta}_1$};

    \node[hollow] (a2) [right =of t1] {$\mbA_2$};
    \node[hollow] (r2) [above =of a2] {$\mbR_2$};
    \node[hollow] (t2) [right =of a2] {$\hat{\Theta}_2$};

    \node[empty] (dots) [right =of t2] {$\ldots$};

    \node[hollow] (at) [right =of dots] {$\mbA_T$};
    \node[hollow] (rt) [above =of at] {$\mbR_T$};
    
    \node[hollow] (x) [above right =of r2] {$\Theta$};
    
    \path[arrow] (x) edge (r1);
    \path[arrow] (a1) edge (r1);
    \path[arrow] (r1) edge (t1);
    \path[arrow] (a1) edge (t1);
    \path[arrow] (t1) edge (a2);

    \path[arrow] (x) edge (r2);
    \path[arrow] (a2) edge (r2);
    \path[arrow] (r2) edge (t2);
    \path[arrow] (a2) edge (t2);
    \path[arrow] (t2) edge (dots);
    
    \path[arrow] (dots) edge (at);
    \path[arrow] (x) edge (rt);
    \path[arrow] (at) edge (rt);
\end{tikzpicture}
\caption{\label{fig:causal-gm}}
\end{subfigure}
\caption{(a) Feedback loops depict the process where the recommendations $\mbA$ affect the ratings $\mbR$. They in turn affect the recommendation $\mbA$ because recommendation algorithms are trained on the collected ratings data. (b) The causal graph of recommender systems  over multiple times steps with feedback loops unrolled.}
\end{figure}
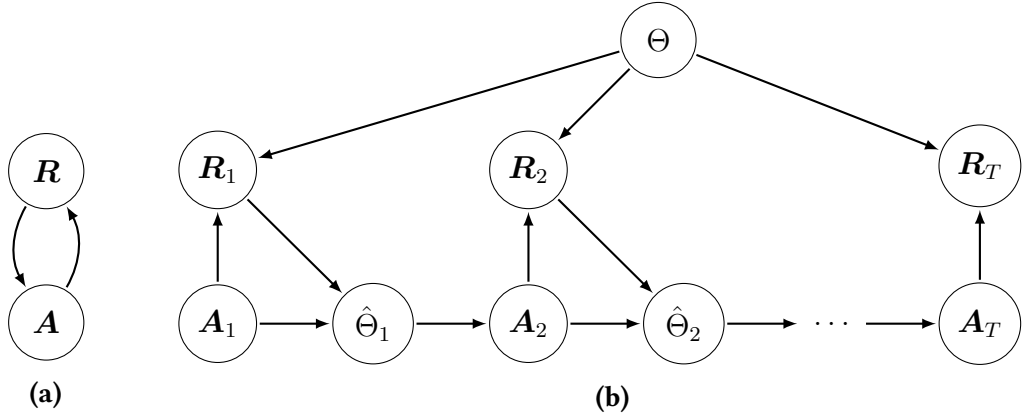

The causal mechanism of recommendation also reveals that the
intervention distributions of recommendations are identifiable from
observational data. To calculate the intervention distributions, it is sufficient to adjust for the recommender system's predictions,
since feedback loops in recommender systems only occur
through this quantity (see \Cref{fig:causal-gm}). Following this
observation, we show how to design an algorithm,
\gls{CAFL}, that estimates intervention distributions in training any loss-minimizing recommendation algorithms. 
\gls{CAFL} enables recommender systems to break feedback loops without
resorting to random recommendations. In particular, it can be applied
to situations where common causal assumptions (e.g.
positivity~\citep{imbens2015causal}) are violated. For example,
\gls{CAFL}  can be applied when a recommender system requires that all
items can only be recommended at most once to each user, which violates the
positivity conditions required by standard causal adjustment methods (e.g. backdoor
adjustment or inverse propensity weighting).

\parhead{Contributions.} The contributions of this work are
three-fold. (1) We formalize the operation of recommender systems over
time as a structural causal model over multiple time steps. (2) We
introduce \gls{CAFL}: a causal adjustment algorithm that can provably
break feedback loops for existing recommendation algorithms.
\gls{CAFL} is easy to implement in existing recommender systems as it
only requires changing the weights of their loss function. (3) Across
multiple simulation environments, we show that \gls{CAFL} not only
corrects the dataset bias caused by feedback loops, but also improves
predictive performance, moreso than prior correction methods.
\gls{CAFL} can also reduce homogenization when feedback loops induce
recommendation homogenization.

\parhead{Related work.} This work is motivated by a recent line of
research that aims to understand the effect of feedback loops in
recommender systems. Using simulations, \citet{schmit2018human} and
\citet{krauth2020offline} have shown that ignoring feedback effects
will negatively impact performance. As recommender systems observe
more data, they are also prone to homogenization and bias
amplification, which researchers often attribute to feedback effects
\citep{chaney2018algorithmic, mansoury2020feedback}. Another related
line of work studies the effects of feedback loops theoretically:
under assumptions of preference drifts, they show that feedback loops
can have undesirable effects on users \citep{jiang2019degenerate,
kalimeris2021preference}. \citet{sharma2015estimating,
hosseinmardi2020evaluating} further illustrate the impacts of
recommender systems and feedback loops using observational data from
large internet platforms.

Feedback loops can induce a sampling bias in the collected user
behavior dataset, so the feedback loop problem is a form of the
\gls{MNAR} problem over multiple time steps
\citep{marlin2009collaborative}. The \gls{MNAR} problem has been
extensively studied in single-step recommender systems: the
recommender system aims to infer missing ratings over a single
timestep using a static dataset. One approach to this problem is to
assume an exposure model between users and items and corrects
the bias such an exposure model would
induce~\citep{sinha2016deconvolving, hernandez2014probabilistic,
liang2016modeling}. Another approach is to use tools from causal
inference to correct this bias under different
assumptions~\citep{schnabel2016recommendations, bonner2018causal,
wang2020causal}. In contrast to these works that focus on single-step
recommender systems, this work studies the setting over multiple time
steps and proposes adjustments based on the special structure of feedback loops.

Correcting feedback effects in recommender system is comparatively
less studied than the \gls{MNAR} problem. Earlier work augments
algorithms with temporal information, but does not model feedback
effects \citep{koren2009collaborative}. More recently,
\citet{sun2019debiasing} combines inverse propensity weighting (IPW)
with active learning to correct for feedback effects, assuming the
same rating model as \citet{liang2016modeling}.
\citet{pan2021correcting} modify their IPW correction to account for
sequential effects. However, their method requires the marginal
probability of an item being observed at each time step. Estimating
these probabilities requires access to observational user data over a
long period of time; the algorithm is thus unable to correct for
feedback effects at the initial period when estimation is performed.
Further, estimating these marginal probabilities is a challenging
task, which compromises the prediction accuracy of algorithms even at
later time steps. In contrast to these works, the \gls{CAFL} algorithm
only requires calculating the probabilities of an item being
recommended conditioned on the history of observed data, which is
available to the recommender system. As a result, \gls{CAFL} is
data-efficient, simple, and does not require strong modeling
assumptions.

Finally, the problem of feedback loops in training machine learning
algorithms is not unique to recommender systems.
\citet{farquhar2021statistical} focus on the problem of active
learning where one collect one data point at each time step; they
propose two unbiased weighting estimators for empirical risk
minimization. \citet{perdomo2020performative} present a general
framework to study the impact of feedback loops in machine learning
predictions. In contrast to these works, the \gls{CAFL} algorithm takes
an explicitly causal view of the feedback loop. It leads to unbiased
weighting estimators that are applicable to recommender systems with
multiple data points acquired in a single time step. The explicit
causal view also enables the use of other causal adjustment strategies
for breaking feedback loops, e.g., backdoor adjustment.

\section{Feedback loops in recommender systems}

\label{sec:feedback}

We describe feedback loops and their consequences in recommender
systems. We focus on \emph{multi-step} recommender systems; these systems
are regularly retrained as they collect more data.

\subsection{Multi-step recommender systems} 

A multi-step recommender system operates over $T$ time steps. At each
time step, it makes recommendations to users, observes their feedback,
and is then retrained on all the data observed so far. 

A multi-step recommender system begins with a set of $U$
new users and $I$ new items. At time step $t=1$, it recommends $N$
items to a random set of users. Denote $\mbA_t$ as the $U\times I$
recommendation matrix, where its $(u,i)$-entry $A_{t,ui}$ is a binary
variable that indicates whether user $u$ was recommended item $i$ at
time $t$. Then, at time $t=1$, we have
$\mbA_{1}\stackrel{iid}{\sim}\mathrm{Multinomial}(n=N,
p=\mathbf{p}_0)$, where $\mathbf{p}_0$ is a matrix of the initial
probabilities of recommendation each item to each user.

After making recommendations, the multi-step recommender then collects
user ratings on the items consumed at this time step. We denote the
data we collect at time $t$ as $\mbR_t$, which is a $U\times I$ rating
matrix; its $(u,i)$-entry $R_{t,ui}$ is user $u$'s rating of item $i$
if the item is consumed at time step $t$, and $R_{t,ui}=0$ otherwise.

Given the user ratings, the multi-step
recommender then infers users' preferences based on the collected
data from this first time step, $\left\{\mbA_1,
\mbR_1\right\}$. Denote $\hat{\Theta}_t$ as the inferred
user preference and item attribute parameters at time $t$, which are
usually minimizers of some loss function. More formally,
the inferred user preference $\hat{\Theta}_t$ can be seen as a maximum
likelihood estimator under some parametric probability model
$P_\Theta$ with parameter $\hat{\Theta}$,
\begin{align}
\label{eq:step-one-matrix-fac}
\hat{\Theta}_1 &= \argmin_\Theta \, \E{\mbA_1}{\mathbb{KL}\left(P\left( \mbR_1\g \mbA_1\right)||P_\Theta\left( \mbR_1\g \mbA_1\right)\right)}\\
&= \argmax_\Theta \E{\mbA_1}{\E{P(\mbR_1\g \mbA_1)}{\log P_\Theta\left(\mbR_1\g \mbA_1\right)}},
\end{align}
where $P\left(\mbR_1\g \mbA_1\right)$ is the (population) conditional
distribution of $\mbR_1$ given $\mbA_1$, from which the collected data $\left\{\mbA_1,
\mbR_1\right\}$ is drawn, and $P_\Theta\left(\mbR_1\g \mbA_1\right)$
is a parametric model we posit. $\mathbb{KL}(\cdot||\cdot)$ denotes
the Kullback–Leibler divergence between the two distributions.

Many existing recommendation algorithms fall into this setup. For
example, assuming a \gls{PROB-MF} model for
ratings~\citep{mnih2008probabilistic} is equivalent to fitting
$\hat{\Theta}$ through a Gaussian linear factor model with maximum
likelihood estimation,
$P_\Theta\left(\mbR_1\g
\mbA_1\right) =\prod_{u=1}^U \prod_{i=1}^I 
\cN\left(R_{1,ui}\g \theta_u^\top \beta_i \cdot A_{1,ui},
\sigma^2\right),$
where the $K\times 1$ user vectors $\theta_u$ and the $K\times 1$ item
vectors $\beta_i$ constitute the parameters $\Theta =
\left((\theta_u)_{u=1}^U, (\beta_i)_{i=1}^I\right)$, and $\sigma^2$ is
assumed known. Similarly, performing weighted matrix
factorization~\citep{hu2008collaborative} and nonnegative
factorization~\citep{gopalan2014content,gopalan2015scalable} also
correspond to fitting parametric probability models using maximum
likelihood~\citep{wang2020causal}.

After fitting a parametric model of user preferences, the recommender system
then recommends items based on the inferred user preferences; that is,
$\mbA_{2,u}{\sim} P_u(\mbA_{2,u} \g \hat{\Theta}_1),
u=1, \ldots, U$, where $P_u(\mbA_{2,u}=1\g \hat{\Theta}_1)$ describes
the probability of each item being recommended to user $u$ based on
the inferred parameters $\hat{\Theta}_1$. This distribution
$P_u(\cdot\g \hat{\Theta})$ captures how recommender systems make
decisions based on user preferences; it is usually specified by
recommender systems \textit{a priori} and can target different goals. For
example, a recommender system may set up $P_u(\cdot\g \hat{\Theta})$
to maximize the potential rating of recommended items, or to increase
user consumption, or to maximize the diversity of the recommendations
without sacrificing users' utility by more than 10\%.

\sloppy
Going from time $t=1$ to time $t=2, \ldots, T$, a \emph{multi-step}
recommender system further updates its inferred preferences, unlike
a \emph{one-step} recommender system which does not update
its inferred user preferences.  At time $t$, it makes new
recommendations $\mbA_t$, collects new data $\left\{\mbA_t,
\mbR_t\right\}$, and updates its parameters $\hat{\Theta}_t$ by
fitting the model to $P(\{\mbR_s\}_{s=1}^{t}\g \{\mbA_s\}_{s=1}^{t})$
using all the data collected up to this point,~$\left\{(\mbA_1,
\mbR_1), (\mbA_2, \mbR_2) \ldots, (\mbA_t, \mbR_t)\right\}.$

But how should a recommender system update its user preference
parameters~$\hat{\Theta}_t$ after the first time step?
In particular, the data collected from
different time steps are not independent. Future recommendations
depend on past ratings because the recommender system tries to
recommend items that users will like, which is inferred from past
ratings. Handling this dependence over time is a core challenge in
developing multi-step recommender systems since naively repeating the optimization problem in Equation~\ref{eq:naive-agg-recsys} would be sub-optimal.

\subsection{Feedback loops in recommender systems} 

To handle the dependence between data points collected at different
time steps, we study their dependency structure. This dependency
structure is often referred to as \emph{feedback loops} in
recommender systems, describing how $\left(\mbA_t,
\mbR_t\right)$---the data collected at time $t$---depends on the data
collected at all previous time steps, $\left\{\left(\mbA_1,
\mbR_1\right),
\ldots, \left(\mbA_{t-1}, \mbR_{t-1}\right)\right\}$.

In more detail, feedback loops refer to the $\mbA\rightarrow
\mbR \rightarrow
\mbA$ loop, going from the recommendation $\mbA_t$, to the rating
$\mbR_t$, and finally back to the recommendation $\mbA_{t+1}$
(\Cref{fig:feedback}). The recommender system begins by
making recommendations $\mbA_t$, these recommendations increase the
probability of the recommended items being rated, hence affecting the
rating we observe $\mbR_t$. The collected ratings $\mbR_t$---together
with all past collected ratings---in turn, affect the recommendation
matrix $\mbA_{t+1}$, because the recommender infers user preferences
from this data and makes recommendations based on this inference.

To learn user preferences from this sequentially collected data, the
most common approach is to aggregate the data from different time
steps \citep{lee2016llorma, sedhain2015autorec, zheng2016neural,
wang2006unifying, rendle:tist2012, steck2019embarrassingly,
ning2011slim}. We fit the probability
model for the first time step  (i.e. \Cref{eq:step-one-matrix-fac}) to
all the data collected up to time $t$, and infer user preferences as
follows:
\begin{align}
\label{eq:naive-agg-recsys}
\hat{\Theta}_t^\mathrm{naive} 
&= \argmax_\Theta \sum_{s=1}^t \E{\mbA_s}{\E{P(\mbR_s\g \mbA_s)}{\log P_\Theta\left(\mbR_s\g \mbA_s\right)}}\\
&= \argmin_\Theta \,  \E{}{\mathbb{KL}\left(\prod_{s=1}^t P\left( \mbR_s\g \mbA_s\right)||\prod_{s=1}^t P_\Theta\left( \mbR_s\g \mbA_s\right) \right)} .
\end{align}
This approach considers a product of the likelihood terms from each
time step, implicitly assuming data $\{\mbA_t, \mbR_t\}$ observed at
different time steps are independently collected. However, they are in
general \emph{not} independent in multi-step recommender systems,
that is,
\begin{align}
\label{eq:time-dependent}
P(\{\mbR_s\}_{s=1}^{t}\g \{\mbA_s\}_{s=1}^{t})  = \prod_{s=1}^t P(\mbR_s\g \mbA_s, \{\mbR_{s'}, \mbA_{s'}\}_{s'=1}^{s-1}) \ne \prod_{s=1}^t P(\mbR_s\g \mbA_s)
\end{align}
since $P(\{\mbA_s,\mbR_s\}_{s=1}^{t})\ne \prod_{s=1}^t P(\mbR_s,
\mbA_s)$. The only exception is when all recommendations $\mbA_s$'s
are completely random, and thus they do not depend on past
observations. By treating dependent data as if they were independent,
$\hat{\Theta}_t^\mathrm{naive}$ is a biased estimator of user
preferences.

Beyond biased estimation, feedback loops are also known to produce a
``rich get richer'' phenomenon, leading to homogenization in
recommendations over time~\citep{chaney2018algorithmic}. As an
example, we contrast two toy movie recommender systems, one with
feedback loops (making recommendations based on
$\hat{\Theta}_t^\mathrm{naive}$ at time $t$) and one without feedback
loops (making random recommendations at time $t$). We then consider their
recommendations to two different users: user A likes both drama and
horror movies, and user B likes both drama and sci-fi movies.

The recommender with feedback loops often experiences recommendation
homogenization over time. For example, suppose it recommends drama
movies to both users at $t=1$ by chance. Both users will rate it
highly because the recommendation aligns with (part of) their
preferences. The recommender then collects these ratings and infers
that both users like drama movies. It thus continues to recommend
drama movies at $t=2$, and again both users will rate it highly.
Continuing with this process, the recommender with feedback loops
will incorrectly infer that the two users have the same preferences,
hence homogenizing user recommendations.

In contrast, the recommender without feedback loops does not
homogenize recommendations. Again suppose it recommends a drama movie
to both users at $t=1$ by chance; both users rate it highly; the
recommender will then infer that both like drama movies, as with the
recommender with feedback loops. At time $t=2$, however, the
recommender will not solely recommend drama movies. Rather, it may
recommend some other movies like a sci-fi movie. In this case, only
user B may rate it highly. With this data from $t=2$, the recommender
will correctly infer that the two users have different user
preferences.

Given these concerns about biased and homogenized recommendations, multi-step recommender systems ought to avoid feedback
loops~\citep{chaney2018algorithmic}. One immediate way to avoid
feedback loops is to adopt a random recommendation mechanism. Such
recommendations are not affected by past ratings and thus are
feedback-free. But the user experience will suffer with random recommendations. How can we
then break the feedback loop without making random recommendations? In
the next sections, we analyze the causal mechanism of feedback loops
and show that a recommender system does not suffer from feedback loops
if it reasons about causal quantities, namely the intervention
distributions of recommendations $\mbA_s$ on ratings $\mbR_s$. This
observation leads us to develop a causal adjustment algorithm that can
provably break feedback loops without resorting to random
recommendations.

\section{Breaking Feedback Loops in Recommendation Systems with Causal
Inference}
\label{sec:break}

We begin with a description of the causal mechanism of feedback loops
in recommender systems. This causal perspective will lead to an
adjustment algorithm that can provably break feedback loops.

\subsection{The causal mechanism of feedback loops} 

To break feedback loops, we first study its causal mechanism by
unrolling it over $t$ time steps.

Begin by writing down the causal graphical
model~\citep{pearl2009causality} of recommender systems in
\Cref{fig:causal-gm}. At time $t=1$, the recommender system begins
with some recommendations $\mbA_1$. These
recommendations causally affect the observed user ratings $\mbR_1$ by
increasing the probability of the recommended items being rated. Both
the recommendations and the observed ratings also affect the inferred
user preference parameters $\hat{\Theta}_1$ since the recommender
optimizes $\hat{\Theta}_1$ based on $\left\{\mbA_1,
\mbR_1\right\}$. This inferred user preference $\hat{\Theta}_1$
then affects $\mbA_2$, i.e. what items are recommended at $t=2$.

The causal structure of $\left\{\mbA_1, \mbR_1,\hat{\Theta}_1, \mbA_2
\right\}$ then repeats itself at each time step $t$. In particular,
the inferred user preferences $\hat{\Theta}_t$ generally depends on
all past recommendations and ratings $\left\{\mbA_{1:(t-1)},
\mbR_{1:(t-1)}\right\}$. Finally, the (unobserved) true user
preferences $\Theta$ affects all ratings across all time steps, hence
there is an arrow from $\Theta$ to all $\left(\mbR_t\right)_{t=1}^T.$

The existence of feedback loops is evident in the causal graph
(\Cref{fig:causal-gm}), where past recommendations and ratings
constantly inform future recommendations. The data $\left\{\mbA_t,
\mbR_t\right\}$ collected at different time steps are thus not
independent, i.e. $P(\{\mbR_s\}_{s=1}^{t}\g \{\mbA_s\}_{s=1}^{t})  \ne
\prod_{s=1}^t P(\mbR_s\g \mbA_s)$ as in \Cref{eq:time-dependent}; they
causally depend on each other, preventing us from fitting a single
model to the data from all time steps. Given the causal graph from
\Cref{fig:causal-gm}, how can we break the feedback loop $\mbA
\rightarrow \mbR \rightarrow \mbA \rightarrow \mbR
\rightarrow \mbA \rightarrow \cdots$ in updating recommendation
algorithms and fitting $\hat{\Theta}_t$?

\subsection{Breaking feedback loops in recommender
systems using causal inference}

To break the feedback loops in \Cref{fig:causal-gm}, we need
to find a distribution $\tilde{P}$ about recommendations and ratings
$\left\{\mbA_s,
\mbR_s\right\}_{s=1}^t$ such that it does achieve independence over
time steps, i.e. $\tilde{P}(\{\mbR_s\}_{s=1}^{t}\g
\{\mbA_s\}_{s=1}^{t})  =
\prod_{s=1}^t \tilde{P}(\mbR_s\g \mbA_s)$. Such a distribution does
not suffer from feedback loops or their resulting model-fitting bias
in \Cref{eq:naive-agg-recsys}; it would allow us to learn user
preference parameters $\hat{\Theta}_t$ from all past time steps.

Causal inference provides a solution to this challenge: the
intervention distribution of recommendations on ratings $P(\mbR_s\g
\mathrm{do}(\mbA_s=\mba))$ achieves this
independence and does not suffer from feedback
loops~\citep{pearl2009causality}. In more detail, $P(\mbR_s\g
\mathrm{do}(\mbA_s=\mba))$ denotes the distribution of $\mbR_s$ under the
intervention of setting $\mbA_s$ to be equal to the value $a$. It does
not suffer from feedback loops because, by definition, a \textit{do}
intervention $\mathrm{do}(\mbA_s=\mba)$ breaks the connection between
the variables being intervened on $\mbA_s$ and its parents
$\hat{\Theta}_{s-1}$ in the causal graph (\Cref{fig:causal-gm}), thus
breaking the $\mbA_s\rightarrow\mbR_s\rightarrow\mbA_{s+1}$ feedback
loop. The following lemma formalizes this argument.

\begin{lemma} \label{lemma:indep-do-intervention}
Assuming the causal graph in \Cref{fig:causal-gm}, we have
\begin{multline}
\label{eq:indep-over-time}
P(\{\mbR_s\}_{s=1}^{t}\g \mathrm{do}(\{\mbA_s\}_{s=1}^{t}=\{\mba_s\}_{s=1}^{t})  =
\prod_{s=1}^t P(\mbR_s\g \mathrm{do}(\mbA_s=\mba_s)),\\ \forall \mba_s \in \{0,1\}^{U\times I}, t\in\{2, \ldots,T\}.
\end{multline}
\end{lemma}
\Cref{lemma:indep-do-intervention} is an immediate consequence of the
\textit{do} calculus~\citep{pearl2009causality}.

Moreover, the intervention distribution is the distribution with the
highest fidelity to the observational data while staying immune from
feedback loops. More precisely, among all distributions $\tilde{P}$
that satisfy independence across time steps
(\Cref{eq:indep-over-time}), the intervention distribution
$\prod_{s=1}^t P(\mbR_s\g
\mathrm{do}(\mbA_s=\mba))$ is the distribution that is closest to
$\tilde{P}\left(\{\mbR_s\}_{s=1}^{t}\g \{\mbA_s\}_{s=1}^{t}\right)$ in
\gls{KL} divergence:
\begin{lemma} \label{lemma:indep-closest-in-KL}
Assuming the causal graph in \Cref{fig:causal-gm}, we have, for all
$\mba_s \in \{0,1\}^{U\times I}$ and $t\in\{2, \ldots,T\}$,
\begin{align}
\label{eq:indep-closest-in-KL}
&\prod_{s=1}^t P(\mbR_s\g \mathrm{do}(\mbA_s=\mba_s))\nonumber\\
&=\argmin_{\tilde{P}\in\mathcal{Q}}
\mathbb{KL}\left(P(\{\mbR_s\}_{s=1}^{t}\g
\{\mbA_s\}_{s=1}^{t}=\{\mba_s\}_{s=1}^{t}) \Big|\Big|
\tilde{P}(\{\mbR_s\}_{s=1}^{t}\g
\{\mbA_s\}_{s=1}^{t}=\{\mba_s\}_{s=1}^{t})\right),
\end{align}
where $\mathcal{Q}=\left\{\tilde{P}: \tilde{P}(\{\mbR_s\}_{s=1}^{t}\g
\{\mbA_s\}_{s=1}^{t}=\{\mba_s\}_{s=1}^{t}) =
\prod_{s=1}^t\tilde{P}(\mbR_s\g \mbA_s=\mba_s)\right\}$ is the set of
all distributions that satisfies the independence relationship in
\Cref{eq:indep-over-time}.
\end{lemma}
\Cref{lemma:indep-closest-in-KL} is an immediate consequence of
Theorem 1 in \citet{wang2019equal}.

Taken together,
\Cref{lemma:indep-do-intervention,lemma:indep-closest-in-KL} imply
that, to avoid feedback loops in recommender system, one should fit
the parametric model $P_\Theta$ to the intervention distributions
$P(\mbR_s\g
\mathrm{do}(\mbA_s=\mba_s))$ instead of the observational
distributions $P(\mbR_s\g \mbA_s=\mba_s)$ in
\Cref{eq:naive-agg-recsys},
\begin{align}
\hat{\Theta}_t^\mathrm{causal} 
&= \argmin_\Theta \,  \E{\{\mbA_s\}_{s=1}^t}{\mathbb{KL}\left(\prod_{s=1}^t P\left( \mbR_s\g \mathrm{do}(\mbA_s=\mbA_s)\right)||\prod_{s=1}^t P_\Theta\left( \mbR_s\g \mbA_s\right) \right)}\\
&= \argmax_\Theta \qquad L^{\mathrm{causal}},\\
&\quad\text{where } \qquad L^{\mathrm{causal}}\triangleq \sum_{s=1}^t \E{\mbA_s}{\E{P(\mbR_s\g \mathrm{do}(\mbA_s=\mbA_s))}{\log P_\Theta\left(\mbR_s\g \mbA_s\right)}}.\label{eq:causal-objective}
\end{align}
We term \Cref{eq:causal-objective} as the \textit{feedback-free causal objective}.
How can we solve the optimization problem in
\Cref{eq:causal-objective} then? How can we estimate the intervention
distributions $P(\mbR_s\g \mathrm{do}(\mbA_s=\mbA_s))$ using the
observational data $\left\{\left(\mbA_1,
\mbR_1\right), \ldots, \left(\mbA_{t}, \mbR_{t}\right)\right\}$? We
discuss these questions in the next section.

\subsection{Inferring user preferences from intervention
distributions}

\label{subsec:ipw-standard}

To estimate the intervention distributions $P(\mbR_s\g
\mathrm{do}(\mbA_s=\mbA_s)$ in recommender systems, we first write
them as a functional of the observational data distribution
$P(\{\mbA_s, \mbR_s, \hat{\Theta}_s\}_{s=1}^t)$. This procedure is
also known as \textit{causal
identification}~\citep{pearl2009causality}. These results will lead to
unbiased estimates of the optimization objectives in
\Cref{eq:causal-objective}, enabling us to infer the parameters
$\hat{\Theta}_t^\mathrm{causal}$ from observational data and perform
recommendation with the inferred parameters.

To identify the intervention distributions $P(\mbR_s\g
\mathrm{do}(\mbA_s=\mbA_s)$, we return to the causal graph
(\Cref{fig:causal-gm}) of multi-step recommender systems. The causal graph implies
that  $P(\mbR_s\g
\mathrm{do}(\mbA_s=\mbA_s)$ is identifiable from the observational
data. Moreover, it is sufficient to adjust for the inferred user
preferences $\hat{\Theta}_{s-1}$ to calculate the intervention
distribution $P(\mbR_s\g
\mathrm{do}(\mbA_s=\mbA_s)$, since the feedback loop
$\mbA\rightarrow\mbR\rightarrow\mbA$ occurs only through
$\hat{\Theta}_{s-1}$ in the causal graph (\Cref{fig:causal-gm}). We
first state the key assumption this argument relies on, namely the
positivity condition, and then state this argument formally.

\begin{assumption}[Positivity, a.k.a. overlap~\citep{imbens2015causal}]
\label{assumption:positivity}
The random variables $\mbA_{s}$ satisfies the positivity condition if,
for all values of $s,\mba_s$ and $\hat{\Theta}_{s-1}$, we have
\begin{align} 
P(\mbA_{s}=\mba_s\g \hat{\Theta}_{s-1}) > 0.
\end{align}
\end{assumption}
Loosely, the positivity condition requires that it must be possible to
recommend any subset of items to any subset of users at any time
steps. Under positivity, we can identify the intervention
distributions as follows.

\begin{lemma}
\label{lemma:backdoor} Assuming the causal graph in
\Cref{fig:causal-gm}. Then under \Cref{assumption:positivity}, we have
\begin{align} P(\mbR_s\g \mathrm{do}(\mbA_s &= \mba_s) ) =\int
P(\mbR_s\g  \mbA_s =
\mba_s,\hat{\Theta}_{s-1})P(\hat{\Theta}_{s-1})\dif
\hat{\Theta}_{s-1}.
\label{eq:backdoor}
\end{align}
\end{lemma}

\Cref{lemma:backdoor} is an immediate consequence of the backdoor
adjustment formula for identifying intervention
distributions~\citep{pearl2009causality}. Plugging \Cref{eq:backdoor}
into \Cref{eq:causal-objective}, \Cref{lemma:backdoor} implies that,
when positivity holds, one can estimate $\hat{\Theta}_t^\mathrm{causal}$
by solving
\begin{align}
\hat{\Theta}_t^\mathrm{causal} 
&= \argmax_\Theta \sum_{s=1}^t \int\int\int P(\mbA_s)P(\mbR_s\g  \mbA_s,\hat{\Theta}_{s-1})P(\hat{\Theta}_{s-1})\log P_\Theta(\mbR_s\g \mbA_s)\dif
\hat{\Theta}_{s-1}\dif \mbA_s \dif \mbR_s\label{eq:raw-causal-objective}\\
&= \argmax_\Theta \sum_{s=1}^t \E{P(\mbA_s,\mbR_s,\hat{\Theta}_{s-1})}{\frac{P(\mbA_s)}{P(\mbA_s\g \hat{\Theta}_{s-1})}\log P_\Theta(\mbR_s\g \mbA_s)},\label{eq:ipw}
\end{align}
where \Cref{eq:ipw} is due to the chain rule
$P(\mbA_s,\mbR_s,\hat{\Theta}_{s-1}) = P(\mbR_s\g
\mbA_s,\hat{\Theta}_{s-1}) \cdot P(\mbA_s\g
\hat{\Theta}_{s-1}) \cdot P(\hat{\Theta}_{s-1})$. 

The expectation in the optimization objective for
$\hat{\Theta}_t^\mathrm{causal}$ (\Cref{eq:ipw}) can be unbiasedly
estimated from observational data. Specifically, \Cref{eq:ipw} implies
that, under positivity, we can solve for
$\hat{\Theta}_{t}^\mathrm{causal}$ via weighted maximum likelihood estimation,
\begin{align}
\hat{\Theta}_{t}^\mathrm{causal}
= \argmax_\Theta \qquad &\hat{L}_{\mathrm{IPW}}^{\mathrm{causal}}(\Theta),\nonumber\\
\text{where  } \qquad
&\hat{L}_{\mathrm{IPW}}^{\mathrm{causal}}(\Theta)\triangleq \sum_{s=1}^t\sum_{\mba_s\in \mathcal{A}}
\frac{\mathbb{1}\{\mbA_s=\mba_s\}}{P(\mbA_s=\mba_s\g
\hat{\Theta}_{s-1})}\cdot \log P_\Theta(\mbR_s\g \mbA_s=\mba_s).\label{eq:ipw-estimator}
\end{align}
The set $\mathcal{A}=\{0,1\}^{U\cdot I}$ denotes the set of all possible
values that $\mbA_s$ can take. The next proposition justifies this estimator.
\begin{prop}[Unbiased estimation of the causal objective (\Cref{eq:causal-objective}) under positivity assumptions]
If positivity (\Cref{assumption:positivity}) holds, then
$\hat{L}_{\mathrm{IPW}}^{\mathrm{causal}}(\Theta)$ is an unbiased
estimator of the causal objective in \Cref{eq:causal-objective},
\begin{align}
\E{}{\hat{L}_{\mathrm{IPW}}^{\mathrm{causal}}(\Theta)} = L^{\mathrm{causal}}(\Theta).
\end{align}
\label{prop:unbiased-causal}
\end{prop}
\Cref{prop:unbiased-causal} is an immediate consequence of
\Cref{eq:ipw,eq:ipw-estimator}. It implies that
$\hat{L}_{\mathrm{IPW}}^{\mathrm{causal}}(\Theta)$ provides a causal
adjustment to the maximum likelihood objective that does not suffer
from feedback loops. It takes the form of the standard
\gls{IPW} estimator in causal inference~\citep{imbens2015causal},
where $\hat{L}_{\mathrm{IPW}}^{\mathrm{causal}}$ is a weighted sum of the one-step
optimization objective for recommendations
(\Cref{eq:step-one-matrix-fac}) with weights being the inverse of the
probabilities $P(\mbA_s\g \hat{\Theta}_{s-1})$.

The weighting estimator
$\hat{L}_{\mathrm{IPW}}^{\mathrm{causal}}(\Theta)$ is different from
the other \gls{IPW} estimators targeting \gls{MNAR} issues in one-step
recommender systems
(e.g.~\citet{schnabel2016recommendations,marlin2009collaborative}):
the weights in $\hat{L}_{\mathrm{IPW}}^{\mathrm{causal}}$ are inverses
of the probabilities given the inferred user preferences at the
previous time step $\hat{\Theta}_{s-1}$; in contrast, the weights in
other \gls{IPW} estimators are often inverses of the probabilities
given user covariates like their demographics or characteristics. This
difference is due to the particular causal structure of feedback loops
in multi-step recommender systems (\Cref{fig:causal-gm}); this
structure is not present in one-step recommender systems.

Taking \Cref{eq:ipw,eq:ipw-estimator} together, we can infer user
preferences $\hat{\Theta}_t$ by fitting matrix factorization models
$P_\Theta$ to intervention distributions, when the positivity
condition holds. For example, we can extend the \gls{PROB-MF} in
\Cref{sec:feedback} from one-step recommender systems to multi-step
recommender systems with feedback loops: we infer user preferences at
each time step by solving
\begin{align*}
&\hat{\Theta}_{t,\mathrm{PROB-MF}}^\mathrm{causal}\\
&= \argmax_\Theta \sum_{s=1}^t\sum_{u=1}^U\sum_{i=1}^I\sum_{a_{s,ui}\in\{0,1\}}
\frac{\mathbb{1}\{A_{s,ui}=a_{s,ui}\}}{P(A_{s,ui}=a_{s,ui}\g
\hat{\Theta}_{s-1})}\cdot \log \cN(R_{s,ui}\g \theta_u^\top \beta_i \cdot A_{s,ui}, \sigma^2)\\
&= \argmax_\Theta \sum_{s=1}^t\sum_{u=1}^U\sum_{i=1}^I
\frac{\mathbb{1}\{A_{s,ui}=1\}}{P(A_{s,ui}=1\g
\hat{\Theta}_{s-1})}\cdot \log \cN(R_{s,ui}\g \theta_u^\top \beta_i , \sigma^2),
\end{align*}
where $\Theta=((\theta_u)_{u=1}^U, (\beta_i)_{i=1}^I)$, and the second
equation is due to $R_{s,ui} = 0$ if and only if $A_{s,ui}=0.$

\subsection{Estimating user preferences under violations of
positivity}

The \gls{IPW} estimator $\hat{L}_{\mathrm{IPW}}^{\mathrm{causal}}$ in
the previous section provides an unbiased estimator of the maximum
likelihood objective for fitting intervention distribution. However,
the validity of $\hat{L}_{\mathrm{IPW}}^{\mathrm{causal}}$ relies on a
core causal assumption, namely the positivity condition
(\Cref{assumption:positivity}) in \Cref{lemma:backdoor}. It requires
that, for all $\mba,s,
\hat{\Theta}_{s-1}$, we have $P(\mbA_{s}=\mba_s\g
\hat{\Theta}_{s-1}) > 0$. In other words, it must be possible to
recommend any subset of items to any subset of users at any time
steps, no matter what the inferred user preferences are.

This positivity condition is often violated in multi-step recommender
systems. For example, multi-step recommender systems often require
that an item cannot be recommended to the same user twice. In such
cases, an item cannot possibly be recommended to a user at a later time
step if it has already been recommended, constituting a violation of
the positivity condition. In such cases, the \gls{IPW} estimator does
not apply since the probability of some recommendation configurations
$P(\mbA_s=\mba_s\g
\hat{\Theta}_{s-1})$ is zero, and thus the inverse probability weight
is infinite.

In this section, we extend the \gls{IPW} estimator
$\hat{L}_{\mathrm{IPW}}^{\mathrm{causal}}$ to settings where
positivity is violated. We construct an unbiased estimator of causal
objective (\Cref{eq:causal-objective}) for these settings. The key
idea is to leverage additional invariance structures of the
intervention distributions $P(R_{s,ui}\g \mathrm{do}(A_{s,ui}))$ over
time to overcome the challenge of positivity violation. Specifically,
we assume that $P(R_{s,ui}\g \mathrm{do}(A_{s,ui}))$ is stationary
over time, and all user-item pairs have a non-zero probability of
recommendation at a non-empty subset of time steps. Then, for $(u,i)$
pairs with probability zero of recommendation at time $t$, we could
form an unbiased estimator for time $t$ using other time steps with
non-zero recommendation probabilities.

\begin{thm}[Unbiased estimation of the causal objective
(\Cref{eq:causal-objective}) under positivity violations] 
\label{thm:ipw-positivity-violation}
Suppose the following assumptions hold:
\begin{enumerate}
  \item There is no interference between users or items at a single
  time step, i.e. recommending an item to a user does not affect the
  ratings of other users or items at the same time step, $P(\mbR_s\g
  \mathrm{do}(\mbA_s=\mba_s)) = \prod_{u=1}^U \prod_{i=1}^I
  P(R_{s,ui}\g \mathrm{do}(A_{s,ui}=a_{s,ui}))$. Moreover, the
  parametric model $P_\Theta$ satisfies a similar factorization.
  $P_\Theta(\mbR_s\g
  \mbA_s=\mba_s) = \prod_{u=1}^U \prod_{i=1}^I
  P_\Theta(R_{s,ui}\g A_{s,ui}=a_{s,ui})$.
  \item The intervention distributions of recommendations on ratings
  are stationary over time, $P(R_{s,ui}\g
  \mathrm{do}(A_{s,ui}=a_{s,ui})) = P(R_{s',ui}\g
  \mathrm{do}(A_{s,ui}=a_{s',ui}))$ for any $s,s'$.
  \item For each user-item pair, there exists some time step when
  there is nonzero probability for this pair to be recommended: for
  each $(u,i)$, there exists some $s\in \{1, \ldots, t\}$ such that
  $P(A_{s,ui}\g \hat{\Theta}_{s-1}) > 0$. 
\end{enumerate}
Then any $\hat{L}_{\mathrm{CAFL}}^{\mathrm{causal}}(\Theta\s \mbc)$ is
an unbiased estimator of the causal objective
(\Cref{eq:causal-objective}): for any $\mbc=(c_1, \ldots, c_t)$ with
$\sum_{s=1}^t c_s=t$, we have
\begin{align}
\E{}{\hat{L}_{\mathrm{CAFL}}^{\mathrm{causal}}(\Theta\s \mbc) } =
L^{\mathrm{causal}}(\Theta),
\end{align}
where
\begin{align}
\label{eq:cafl}
&\hat{L}_{\mathrm{CAFL}}^{\mathrm{causal}}(\Theta\s \mbc) \nonumber\\
&\triangleq \sum_{s=1}^t c_s \left[ \sum_{\substack{u,i: \\P(A_{s,ui}=\\a_{s,ui}\g \hat{\Theta}_{s-1})=0}} \sum_{a_{s,ui}\in\{0,1\}}\hat{L}_{s,u,i}^{\mathrm{unobs}}(\Theta\s a_{s,ui})+ \sum_{\substack{u,i: \\P(A_{s,ui}=\\a_{s,ui}\g \hat{\Theta}_{s-1})>0}} \sum_{a_{s,ui}\in\{0,1\}}\hat{L}_{s,u,i}^\mathrm{obs}(\Theta\s a_{s,ui})\right],
\end{align}
with 
\begin{align*}
\hat{L}_{s,u,i}^\mathrm{unobs}(\Theta\s a_{s,ui}) &\triangleq \frac{\sum_{r: P(A_{r,ui}=a_{s,ui}\g \hat{\Theta}_{r-1})>0} \mathbb{1}\left\{A_{r,ui}=a_{s,ui}\right\}\cdot\log P_\Theta\left(R_{r,ui}\g A_{r,ui}=a_{s,ui}\right) }{\sum_{r: P(A_{r,ui}=a_{s,ui}\g \hat{\Theta}_{r-1})>0} \mathbb{1}\left\{A_{r,ui}=a_{s,ui}\right\}},\\
\hat{L}_{s,u,i}^\mathrm{obs}(\Theta\s a_{s,ui}) &\triangleq \frac{\mathbb{1}\left\{A_{s,ui}=a_{s,ui}\right\}\cdot \log P_\Theta\left(R_{s,ui}\g A_{s,ui}=a_{s,ui}\right) }{P\left(A_{s,ui}=a_{s,ui}\g \hat{\Theta}_{s-1}\right)}.
\end{align*}
\end{thm}

The proof of \Cref{thm:ipw-positivity-violation} is in
\Cref{app:causal-proof}. Loosely,
$\hat{L}_{\mathrm{CAFL}}^{\mathrm{causal}}(\Theta\s \mbc)$ treats the
entries where positivity holds in $\hat{L}_{s,u,i}^\mathrm{obs}$ and
the other entries where positivity is violated in
$\hat{L}_{s,u,i}^\mathrm{unobs}$. The term
$\hat{L}_{s,u,i}^\mathrm{obs}$ is the standard \gls{IPW} estimator as
in \Cref{prop:unbiased-causal}. The term
$\hat{L}_{s,u,i}^\mathrm{unobs}$ leverages the stationary assumption
on intervention distributions to form an unbiased estimate for entries
with positivity violations, namely the empirical average of the likelihood term at other time steps.

\Cref{thm:ipw-positivity-violation} suggests
that, to avoid feedback loops in recommender systems, we should make
recommendations at each time step by solving for
\begin{align}
\label{eq:cafl-adjust}
\hat{\Theta}_t^{\mathrm{causal}} = \argmax \qquad \hat{L}_{\mathrm{CAFL}}^{\mathrm{causal}}(\Theta\s \mbc)
\end{align}
at each time step $t$.

In the special case where the recommender systems do not allow the
same item to be recommended twice, we can instantiate
\Cref{thm:ipw-positivity-violation} in the following corollary.

\begin{corollary}
\label{corollary:special-case}
When (1) only one user-item pair is recommended at each time step, and
no item can be recommended twice to the same user, (2) all assumptions
of \Cref{thm:ipw-positivity-violation} hold, we have
\begin{align}
\label{eq:special-case}
&\hat{L}_{\mathrm{CAFL}}^{\mathrm{causal}}(\Theta\s \mbc)\nonumber\\
&= \sum_{u, i}\sum_{s=1}^t c_s  \left[ \sum_{r < s} \mathbb{1}\left\{A_{r,ui} = 1\right\}\log P_\Theta\left(R_{r,ui}\g A_{r,ui}\right) + \frac{\mathbb{1}\left\{A_{s,ui} = 1\right\}\log P_\Theta\left(R_{s,ui}\g A_{s,ui}\right) }{P\left(A_{s,ui}\g \hat{\Theta}_{s-1}\right)}\right]\nonumber\\  
&\qquad + \mathrm{constant} 
\end{align}
\end{corollary}
\Cref{corollary:special-case} is an immediate consequence of
\Cref{thm:ipw-positivity-violation}. The first term in
\Cref{eq:special-case} considers the terms where $A_{s,ui}=1$,
and the constant term absorbs the $A_{s,ui}=0$ terms because
$R_{s,ui}=0$ if and only if $A_{s,ui}=0$.

\parhead{Choosing the constants $\mbc$.} A final challenge is to choose the constants $\mbc$ in the unbiased estimators~(\Cref{eq:cafl,eq:special-case}), since any
constant $\mbc$ with $\sum_{s=1}^t c_s=t$ will lead to an unbiased estimator
of \Cref{eq:causal-objective}. A common choice is to choose $\mbc$
such that the expectation of the weights in front of each term $\log
P_\Theta(R_{s,ui}\g A_{s,ui})$ is the same, since all $(R_{s,ui},
A_{s,ui})$ pairs shall be similarly informative for $\Theta$. 

In the special case of \Cref{corollary:special-case}, one can obtain
the $\mbc$ vector by calculating thet expected weights in front of
each term. In more detail, the expected weight of the $s$-th term $\log
P_\Theta(R_{s,ui}\g A_{s,ui})$ is $\sum_{s'=s+1}^t c_{s'} + c_s
\cdot(UI-s+1) / UI$. The reason is that the $s$th term does not
contribute to $\hat{L}_{\mathrm{CAFL}}^{\mathrm{causal}}(\Theta\s
\mbc)$ in the first $s'=1, \ldots, s-1$ time steps, i.e. before the
first occurrence where $A_{s,ui}=1$. It then contributes with weight
$c_s / P(A_{s,ui}\g
\hat{\Theta}_{s-1})$ at the $s$th time step, where $P(A_{s,ui}\g
\hat{\Theta}_{s-1})$ has an expectation of $(UI-s+1) / UI$ since
$UI-s+1$ items out of the total $UI$ items remains to have nonzero
probability of being recommended. Finally, it contributes weight
$c_s'$ for all future time steps $s'=s+1,\ldots, t$ due to the first
term of \Cref{eq:special-case}. This calculation leads to $c_s= (UI
(UI-t)) / ((UI-s)(UI-s+1))$, which repeats the calculation in
Appendix B.4. of \citet{farquhar2021statistical}.

\parhead{Intuitive understanding of the weighting estimator.} Given the weighting estimator in \Cref{thm:ipw-positivity-violation}, we next develop some intuitive understanding of the weights in the
special case of \Cref{corollary:special-case}. Focusing on the choice
of $c_s= (UI (UI-t)) / ((UI-s)(UI-s+1))$ above, the weight $W_{ui}$ of
each log-likelihood term $\log P_\Theta\left(R_{s,ui}\g
A_{s,ui}=1\right)$ is
\begin{equation}
    W_{ui} = \underbrace{\frac{UI}{UI - s}}_{\text{Normalization}}\left(\underbrace{t - s}_{\text{Fix 1}} + \underbrace{\frac{UI - t}{UI - s + 1}}_{\text{Fix 2}}\underbrace{P\left(A_{s,ui}=a\g\hat{\Theta}_{t-1}\right)^{-1}}_{\text{IPW Weight}}\right),
    \label{eq:weight-intuition}
\end{equation}
where $s$ is the time at which that user-item pair $(u,i)$ was
observed and $t$ is the current time step. To understand these weights
intuitively, we first notice that, in \Cref{eq:weight-intuition}, the
\gls{IPW} weight ensures that likely $(u,i)$-pairs will be down-weighted
while unlikely pairs will be up-weighted. This weighting scheme mimics
a uniformly sampled distribution over the set of remaining items, as
with standard \gls{IPW} weighting. It ensures
$\mathbb{E}\left[W_{ui}P\left(A_{t,ui}\g
\hat{\Theta}_{t-1}\right)\right]$ is constant across all unobserved
$(u,i)$-pairs at time $t$.

Beyond the \gls{IPW} weighting, $W_{ui}$ also exerts several fixes. Fix
1 upweights the observed $(u,i)$-pair since it has not had a chance at
being recommended for $t - s$ time steps, this mimics a uniformly
sampled distribution over the set of \textit{all} items. In
combination with the \gls{IPW} weight, Fix 1 ensures
$\mathbb{E}\left[W_{ui}P\left(A_{t,ui}\g
\hat{\Theta}_{t-1}\right)\right]$ is constant across all $(u,i)$-pairs
both observed and unobserved. Next, Fix 2 accounts for the fact that
if a $(u,i)$-pair is recommended earlier on, it likely had a smaller
chance of being recommended at that time step than if it were
recommended at a later time step. Hence, \gls{IPW} weights will tend
to be larger for $(u,i)$-pairs recommended early, which implies their
\gls{IPW} weight should be downweighted as time progresses. By
rewriting Fix 2 as $\frac{1}{UI-s+1}\left(\frac{1}{UI-t}\right)^{-1}$,
we observe that it is equal to the \gls{IPW} weight of a random
recommender at time step $t$ multiplied by the inverse of the IPW
weight of a random recommender at time step $s - 1$. In this sense, we
are effectively canceling out the effect of the sample space's size
on the \gls{IPW} weight at time $s - 1$ and replacing it with the
effect of the sample space's size at time $t$.

\parhead{Applicability of \gls{CAFL} to general probability models and time-varying user preferences.} Zooming out from the special case of
\Cref{corollary:special-case}, we note that the \gls{CAFL} algorithm
can be applied to any common parametric probability model originally
developed as a one-step recommender system. The reason is that most
recommendation models satisfy the constraints (especially
the first assumption on $P_\Theta(\cdot)$) in
\Cref{thm:ipw-positivity-violation}. While we mainly illustrated the
adjustment with \gls{PROB-MF} in \Cref{subsec:ipw-standard}, 
\gls{CAFL} can be applied to general matrix factorization models,
including weighted matrix factorization~\citep{hu2008collaborative},
Poisson matrix factorization~\citep{gopalan2014content}, variational
autoencoders~\citep{liang2018variational}.

The \gls{CAFL} algorithm can also be extended to accommodate time-varying user
preferences. We simply replace the time-invariant parametric model
$P_\Theta$ with a time-varying one $P^t_{\Theta}$, indicating the user
behavior patterns at time $t$. To infer $P^t_{\Theta}$, we can extend
the parametric model to take in time as a parameter, e.g.
$P^t_{\Theta}(\mbR_t\g \mbA_t; t)$. For example, one can extend
\gls{PROB-MF} to its time-varying counterpart with $R_{s,ui}\sim
\cN((g(t, \theta_u)^\top \beta_i \cdot A_{s,ui}, \sigma^2)$ for some
parametric function $g$ (e.g., a neural network). Such a time-varying
parametric model, together with \gls{CAFL}, will enable us to handle
time-varying user preferences in the presence of feedback loops.

\parhead{Connections between \gls{CAFL} and existing \gls{IPW} estimators.} We conclude the section by discussing the connections between
\gls{CAFL} and other similar-looking \gls{IPW} estimators. We begin
with the connection between \gls{CAFL} and existing \gls{IPW} weigting
estimators in one-step recommender
systems~\citep{schnabel2016recommendations,marlin2009collaborative}.
The $\hat{L}_{\mathrm{cafl}}(\Theta)$ objective differs from these
standard \gls{IPW} estimators in one-step recommender systems, because
these latter estimators often rely on adjusting for the probability
of recommendation given external user characteristics or covariates.
They are different from \gls{CAFL}, which adjusts for the
probability given inferred user preferences from the previous time
step. This adjustment of \gls{CAFL}, in particular its sufficiency for breaking
feedback loops, is unique to the multi-step recommender systems we
consider here. Moreover, existing estimators always assume positivity
(\Cref{assumption:positivity}); they cannot provide unbiased
estimation in multi-step recommendation settings where some variables
may not be freely manipulated at certain time steps.

Finally, in the special case of \Cref{corollary:special-case}, the
$\hat{L}_{\mathrm{cafl}}(\Theta)$ objective coincides with the
$R_{\mathrm{LURE}}$ estimator of \citet{farquhar2021statistical}.
However, the $\hat{L}_{\mathrm{cafl}}(\Theta)$ in
\Cref{thm:ipw-positivity-violation} can be applied to general
statistical estimation and causal inference settings when feedback
loops are present. For example, it extends to settings where multiple
data points are acquired at each time step and/or where data points
acquired need not be distinct across time steps. The derivations of
the two estimators are also complementary to each other:
$R_{\mathrm{LURE}}$ is constructed by finding weights that do not
depend on the time at which data points are collected. In contrast,
\Cref{eq:special-case} is derived from an explicitly causal
perspective and finding the optimal distribution that does not suffer
from feedback loops.

\section{Empirical Studies}
\label{sec:empirical}
In this section, we evaluate the proposed \gls{CAFL} algorithm using two environments from the RecLab simulation framework \citep{krauth2020offline}. We show that \gls{CAFL} mitigates negative feedback effects:
\begin{enumerate}
    \item \gls{CAFL} corrects the dataset bias caused by feedback loops and improves the predictive performance and recommendation quality of our model. 
    \item \gls{CAFL} reduces homogenization when feedback loops induce homogenization.
    \item In settings where feedback loops do not cause homogenization, we show that the behavior of \gls{CAFL} tracks random sampling, suggesting that \gls{CAFL} breaks feedback loops.
\end{enumerate}

Furthermore, we compare \gls{CAFL} in the experiment proposed by \citet{pan2021correcting} in Section~6.3 of their paper. We show that \gls{CAFL} significantly outperforms both the correction method proposed by \citet{pan2021correcting}, Poisson factorization \citep{gopalan2015scalable}, a popularity-based correction, and uncorrected matrix factorization.

\subsection{Evaluation of feedback effects in recommender systems}

In this section we detail the experimental setup in Sections~\ref{sec:expt-quality} and \ref{sec:expt-homogenization}. We outline the experimental setup comparing \gls{CAFL} to prior work in Section~\ref{sec:expt-comparison}.

\parhead{Metrics of feedback effects.} We begin by defining the metric we use for measuring feedback effects.
\begin{defn}
 Let $\widetilde{\mbA}_t$ be the recommendation matrix from a random recommender and let $\widetilde{\mbR}_t$ be the corresponding rating matrix. Furthermore, let $\widehat{\mbA}_t \sim P(\widetilde{\Theta}_{t - 1})$ be the recommendation matrix where $\widetilde{\Theta}_{t-1}$ are the parameters derived by observing the ``shadow'' randomly sampled dataset $\{(\widetilde{\mbA}_1, \widetilde{\mbR}_1), (\widetilde{\mbA}_2, \widetilde{\mbR}_2), \ldots, (\widetilde{\mbA}_{t-1}, \widetilde{\mbR}_{t-1})\}$.
 Then the effect of feedback at time $t$ with respect to some metric $\mathcal{M}$ is defined as:
 \[\mathcal{M}\left(\mbR_t\right) - \mathcal{M}\left(\widehat{\mbR}_t\right).\]
\end{defn}
This definition states that the effect of feedback is the difference in performance between a model that observes the ratings of the items it recommends, and a model that observes the ratings of items drawn at random. This definition allows us to differentiate between true feedback effects and phenomena caused by confounding factors, such as the inductive bias of matrix factorization models.

While $\widehat{\mbR}_t$ is not usually observable and must be approximated, it can easily be computed in simulated environments. Therefore, we report $\mathcal{M}\left(\widehat{\mbR}_t\right)$ in all the experiments to quantify feedback effects.

\parhead{Simulation environments.} The two simulated environments we consider are beta-rank-v1 and ml100k-v1. Beta-rank-v1 is an implementation of the environment developed by \citet{chaney2018algorithmic}, it is of significance since it was used to demonstrate that recommender systems under feedback loops homogenize user interactions. Ml100k-v1 represents each user and item as a 100-dimensional latent vector. These latent vectors were generated by fitting a matrix factorization model to the MovieLens 100K dataset \citep{harper2015movielens}. Ml100k-v1 allows us to evaluate the algorithms in a simulation initialized with a real-world dataset.

Across all experiments, we use matrix factorization trained using alternating least squares as the parametric model $P_\Theta$~\citep{bell2007scalable}. We implement \gls{CAFL} in this setting by running weighted least squares, where the weight for each observed rating $R_{t, ui}$ is as defined in Equation~\ref{eq:special-case}. We repeat each experiment $10$ times and report results averaged across all runs.

\parhead{Competing methods.} We compare \gls{CAFL} with two other algorithms: (1) matrix factorization re-trained on the newly collected data at each time step; (2) matrix factorization trained on uniformly sampled unrecommended items at each time step. The first algorithm is the usual multi-step recommendation algorithm that suffers from feedback loops; it is a baseline algorithm on which \gls{CAFL} performs causal adjustment and improves upon. The second uniform sampling approach does not suffer from feedback loops because it observes randomly sampled data (although makes non-random recommendations). If the performance of an algorithm tracks uniform sampling, then it suggests that it does not suffer from feedback loops.

\subsection{\gls{CAFL} improves recommendation quality}\label{sec:expt-quality}
In this section, we evaluate how \gls{CAFL} impacts the model's accuracy over time. To evaluate the algorithm in an unbiased manner, we create a test set by randomly sampling user-item pairs. User-item pairs in the test set can not be recommended during the run.

\parhead{Evaluation metrics for recommendation quality.}
We evaluate each algorithm using root mean squared error (RMSE), which measures predictive accuracy, and normalized discounted cumulative gain (NDCG), which measures ranking accuracy and places a heavier emphasis on higher rankings. We report RMSE and NDCG with respect to the held-out test set.

\begin{figure}[!htbp]
    \centering
    \includegraphics[width=0.49\textwidth]{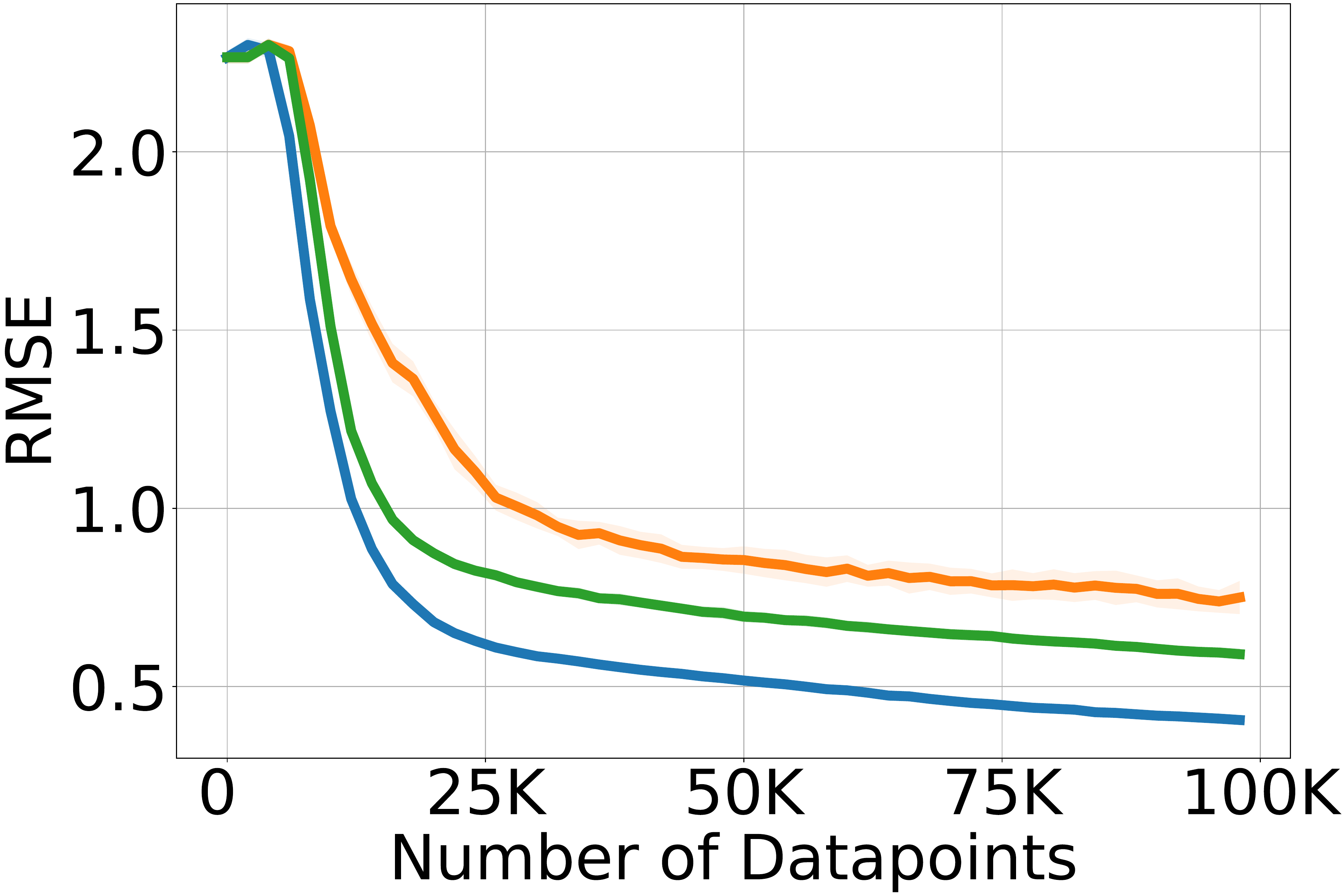}
    \includegraphics[width=0.49\textwidth]{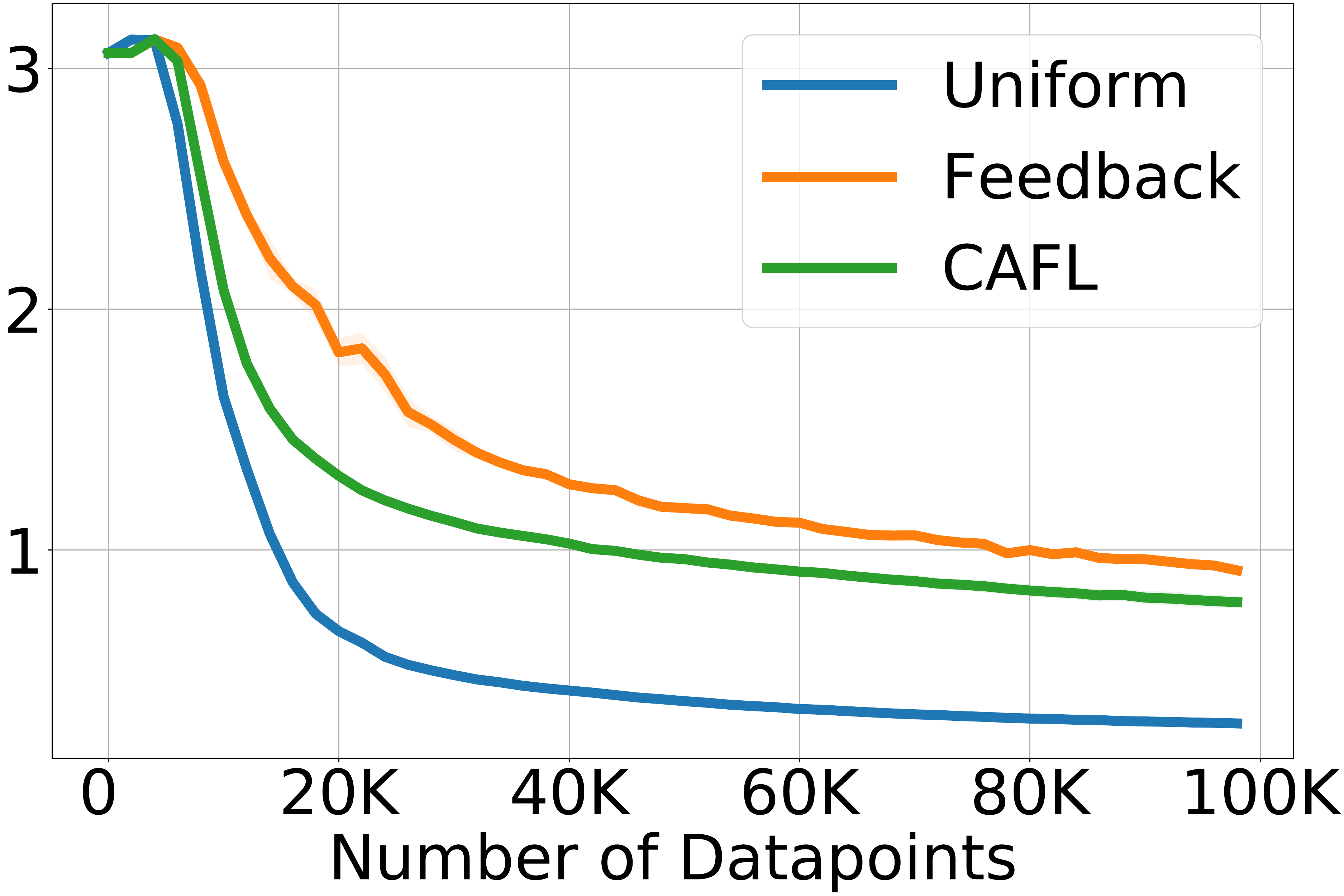}
    \caption{The mean RMSE of the models in the beta-rank-v1 (left) and ml-100k-v1 (right) environments averaged across $10$ runs. Shaded areas indicate $95\%$ confidence intervals. RMSE was evaluated with respect to a randomly sampled test set of size 100,000. Both \gls{CAFL} and Feedback observe their own recommendations, while Uniform observes randomly chosen user-item pairs. Lower RMSE is better.}
    \label{fig:rmse}
\end{figure}
\begin{figure}[!htbp]
    \centering
    \includegraphics[width=0.49\textwidth]{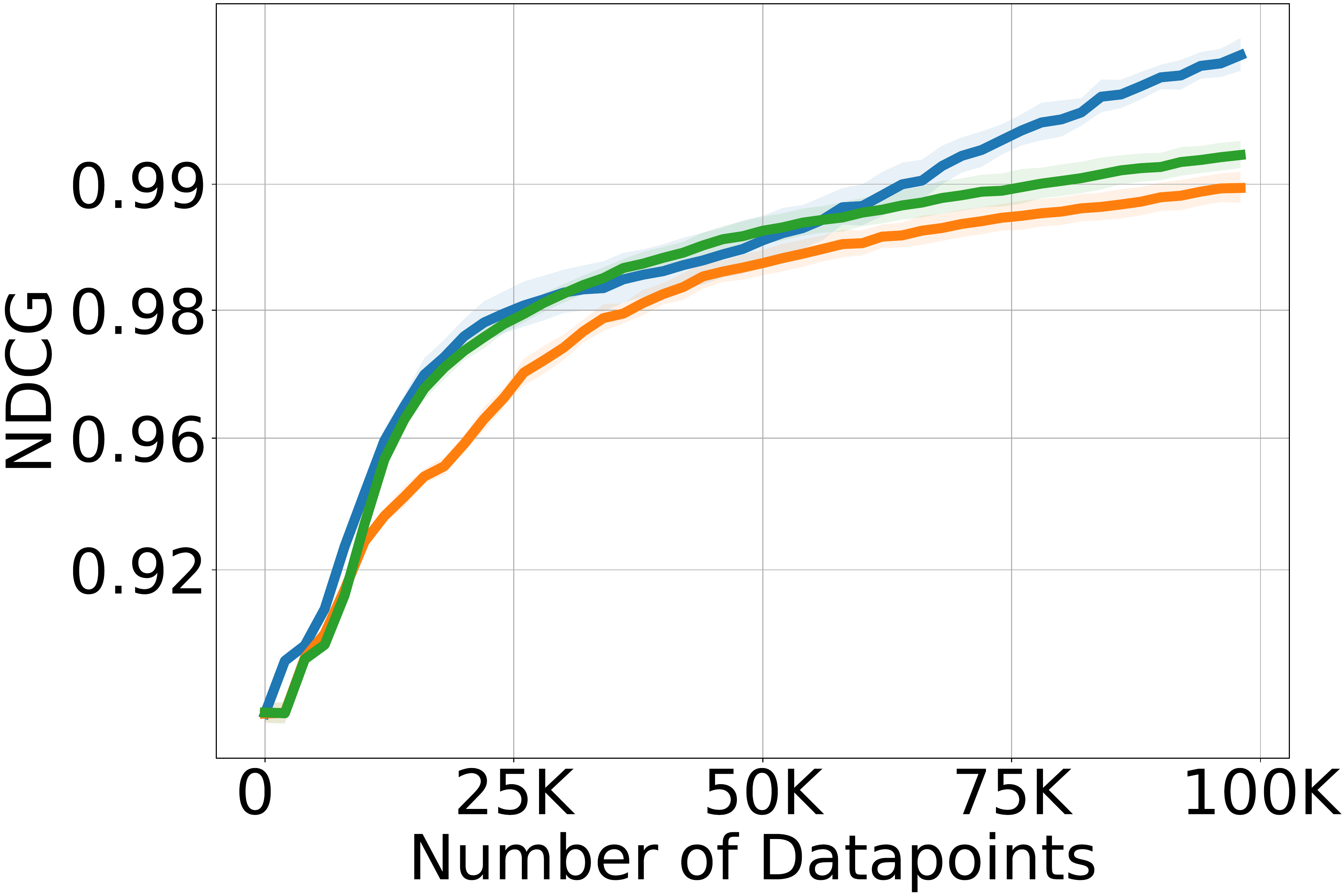}
    \includegraphics[width=0.49\textwidth]{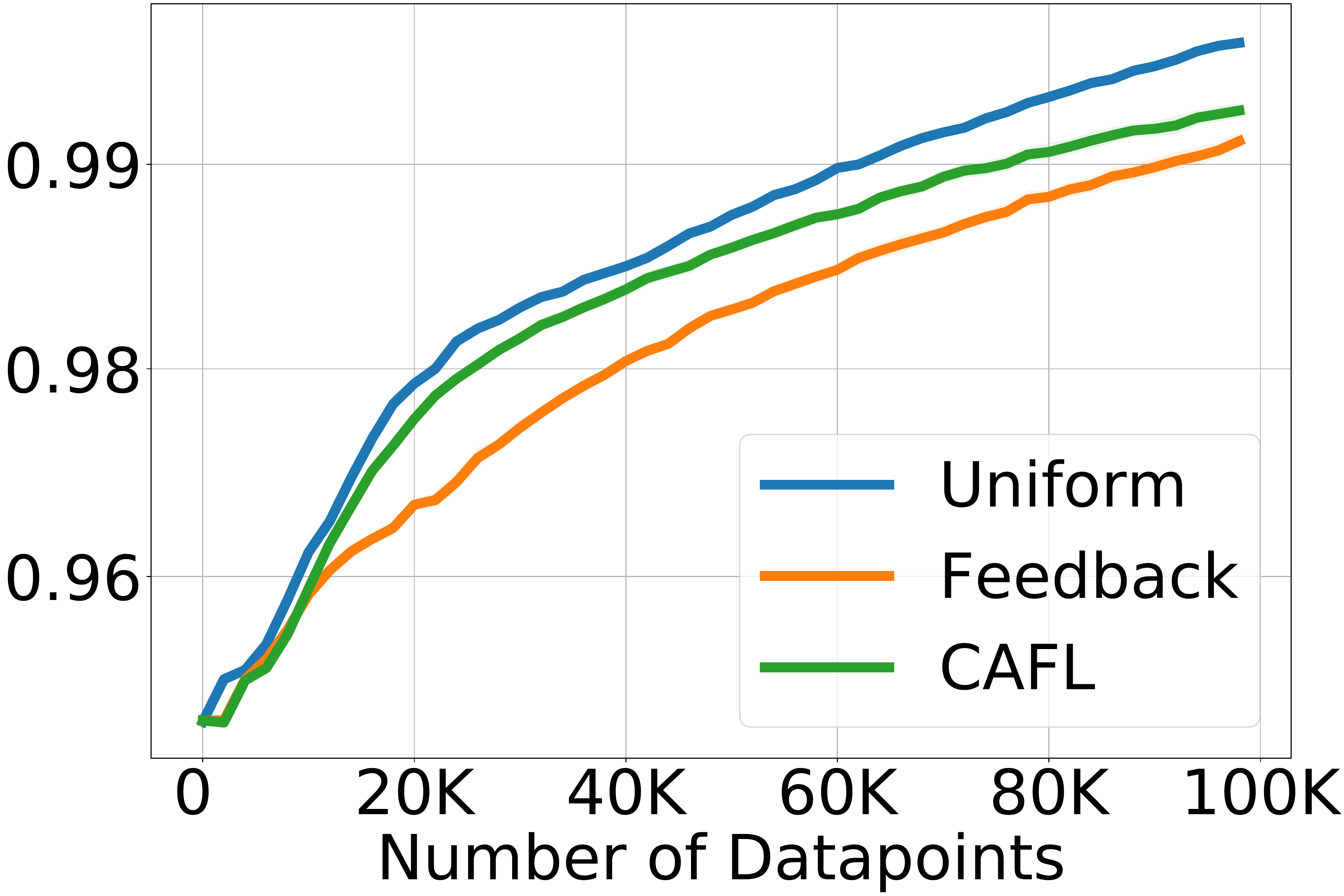}
    \caption{Left: The mean NDCG of the models averaged across all users and across $10$ runs in the beta-rank-v1 (left) and ml-100k-v1 (right) environments. Shaded areas indicate $95\%$ confidence intervals. NDCG was evaluated with respect to a randomly sampled test set of size 100,000. Both \gls{CAFL} and Feedback observe their own recommendations, while Uniform observes randomly chosen user-item pairs. We use a logit scale for the Y-axis for readability. Higher NDCG is better.}
    \label{fig:ndcg}
\end{figure}

\parhead{Results.} As shown in \Cref{fig:rmse,fig:ndcg}, \gls{CAFL} increases the model's predictive (RMSE) and ranking (NDCG) accuracy, when compared to the uncorrected version (Feedback).We note that the model that observes uniformly chosen datapoints (Uniform) still outperforms \gls{CAFL} in most cases. This is expected since the \gls{CAFL} correction is attempting to use the observed feedback data to approximate the empirical risk that Uniform observes. Uniform effectively observes more datapoints than \gls{CAFL} at any given timestep. 

\subsection{\gls{CAFL}, feedback loops, and homogenization}\label{sec:expt-homogenization}
Recommendation systems and their feedback loops have been shown to homogenize the set of items that users will observe beyond what is necessary to achieve optimal utility \citep{chaney2018algorithmic}. This is troublesome since it implies algorithmic minutia may have an undeservedly large impact on the popularity of different items. 

Here we evaluate the homogenization effect of uniform sampling, \gls{CAFL}, and the vanilla recommender with feedback loops. We show that \gls{CAFL} reduces homogenization when feedback loops induce homogenization. In settings where feedback loops do not induce homogenization (i.e. when feedback loops induce the same or less homogenization than uniform sampling), we show that the behavior of \gls{CAFL} tracks random sampling, suggesting that \gls{CAFL} breaks feedback loops in those settings too.

\parhead{Evaluation metrics for homogenization.} We define homogenization as the mean similarity between every pair of users' recommended items, for which we use the Jaccard coefficient as the measure of similarity between two different users $u_1$ and $u_2$:
\[S(u_1, u_2) = \frac{\sum_{i} \mbA_{t, u_1i} \wedge \mbA_{t, u_2i}}{\sum_{i}\mbA_{t, u_1i} \vee \mbA_{t, u_2i}}.\]

\begin{figure}[!htbp]
    \centering
    \includegraphics[width=0.49\textwidth]{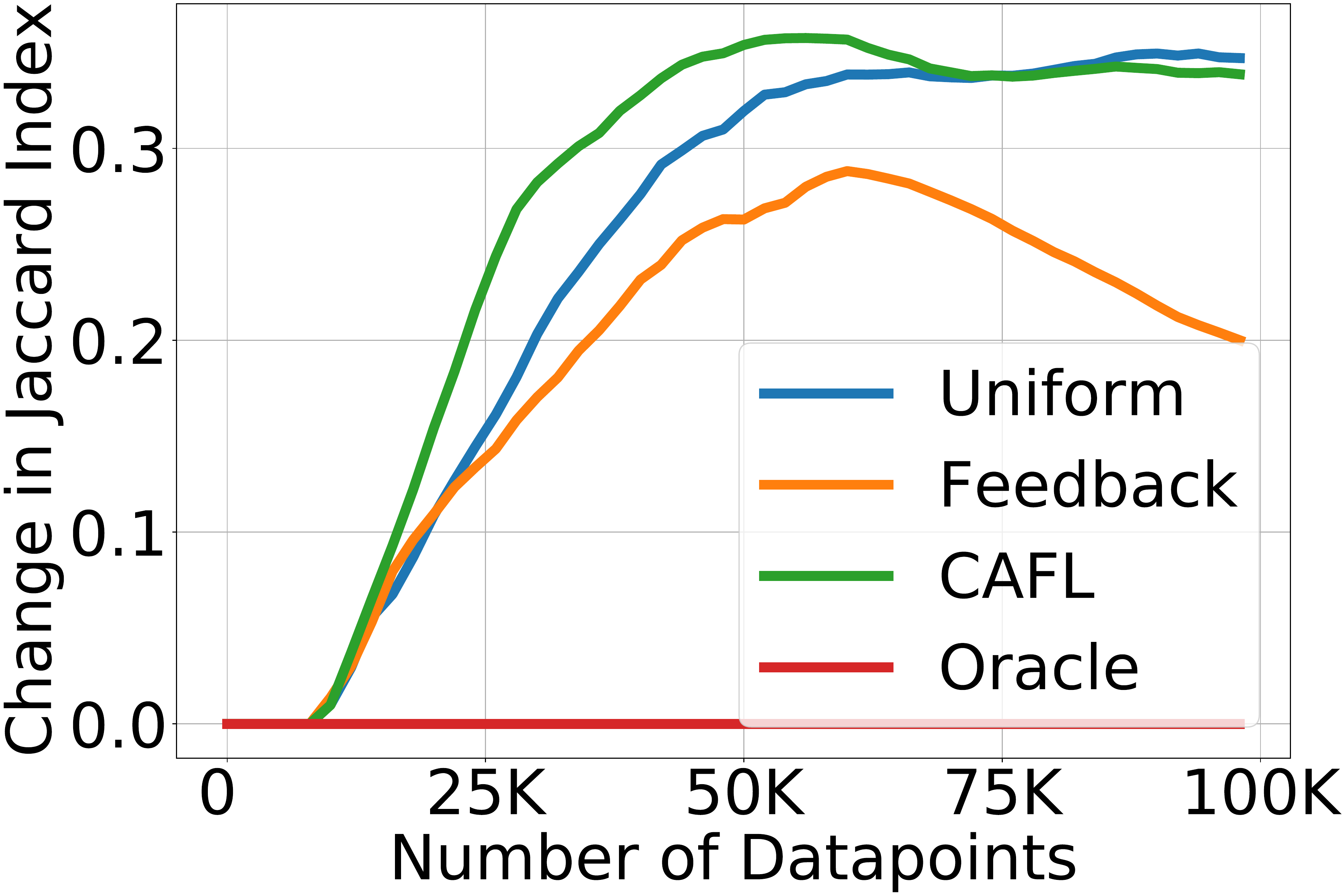}
    \includegraphics[width=0.49\textwidth]{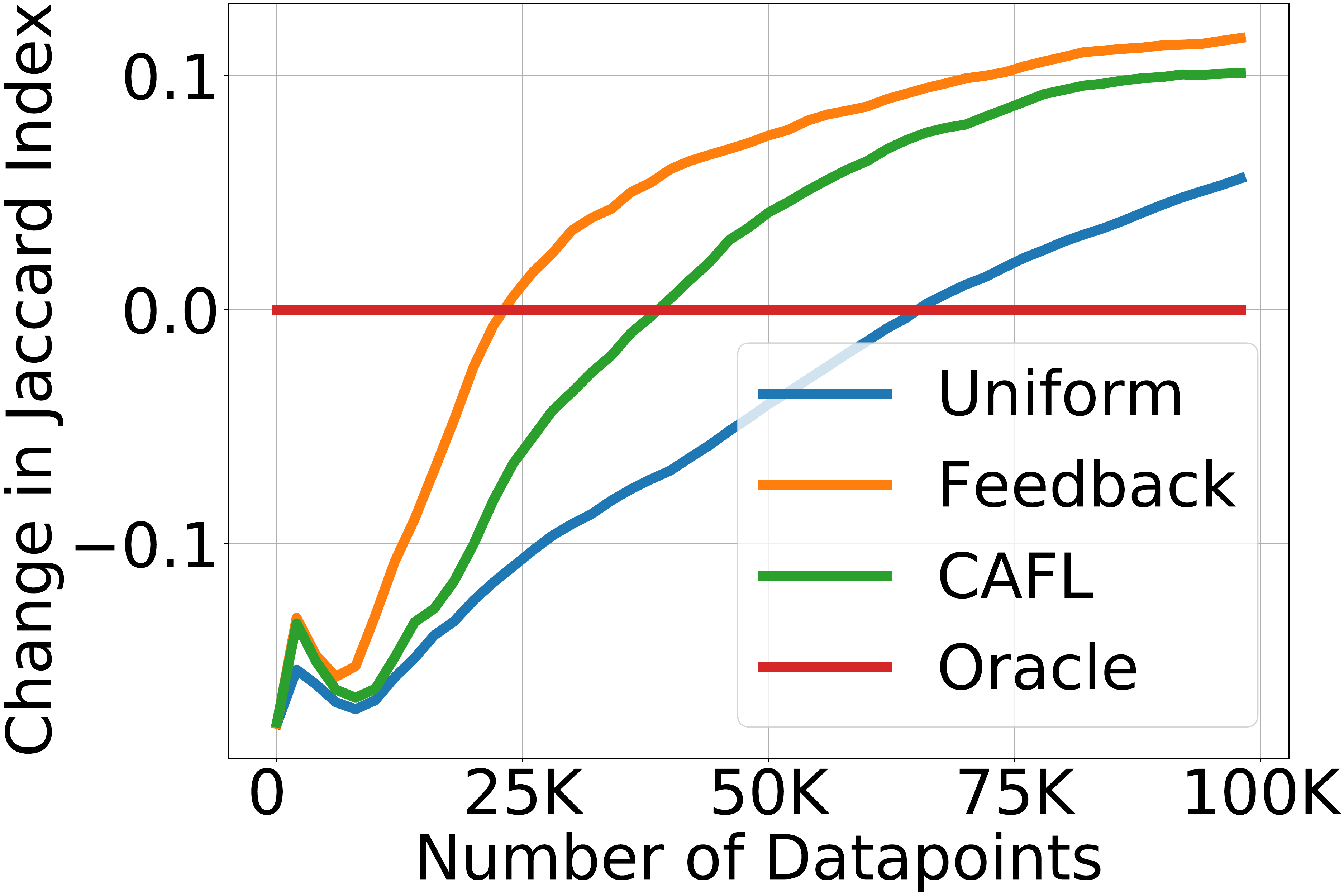}
    \caption{The mean Jaccard coefficient between the set of recommended items of each user-item pair at each timestep minus the Jaccard coefficient of an oracle recommender system on the beta-rank-v1 (left) and ml100k-v1 (right) environments. Both \gls{CAFL} and Feedback observe their own recommendations, while Uniform observes randomly chosen user-item pairs. Lower change in Jaccard index is better.}
    \label{fig:jaccard}
\end{figure}

\parhead{Results.} When feedback loops increase homogenization, \gls{CAFL} successfully mitigates homogenization. The right plot of Figure~\ref{fig:jaccard} shows the Jaccard index over time for the ml-100k-v1 environment. In this setting, feedback effects cause the uncorrected recommender system to further homogenize the user experience when compared to a recommender system that observes uniformly sampled data. \gls{CAFL} is able to reduce homogenization in this setting. We note that this outcome is not self-evident. In particular, the \gls{CAFL} correction only leads to a more accurate empirical risk estimate and does not explicitly consider homogenization.

Turning to the beta-rank-v1 homogenization results, we observe that \gls{CAFL} is unable to reduce homogenization when it is not caused by feedback effects. As shown in the left plot of Figure~\ref{fig:jaccard}, \gls{CAFL} increased homogenization in this setting when compared to the uncorrected feedback recommender. Surprisingly, the uniform recommender also leads to higher homogenization. This suggests that homogenization is not always caused by feedback effects, since we would otherwise expect the feedback recommender to have the highest homogenization if that were the case. In fact, these results suggest that feedback can sometimes reduce homogenization.

\subsection{Comparison with Prior Work}\label{sec:expt-comparison}
We replicated the experimental setup of \citet{pan2021correcting} to compare \gls{CAFL} with prior correction methods. We evaluate \gls{CAFL} on a variation of the simulated environment first proposed by \citet{chaney2018algorithmic}. In this setup if user $u$ interacts with item $i$ at time $t$ we have
\begin{align*}
    R_{t,ui} &\sim \mathrm{Beta}'\left(\theta_{u}^\top \beta_{i}\right)
\end{align*}
where $\mathrm{Beta}'(\mu)$ is a reparametrized beta distribution with variance $\sigma^2 = 0.01$ that is equivalent to $Beta(a, b)$ where $a = \left(\frac{1 - \mu}{\sigma^2} - \frac{1}{\mu}\right)\mu$ and $b = a\left(\frac{1}{\mu} - 1\right)$. The latent user and item vectors have distribution 
\begin{align*}
    \theta_u \sim \mathrm{Dirichlet}(\mu_\theta),&\quad \mu_\theta \sim \mathrm{Dirichlet}(\mathbf{20})\\
    \beta_i \sim \mathrm{Dirichlet}(\mu_\beta),&\quad \mu_\beta \sim \mathrm{Dirichlet}(\mathbf{100}).
\end{align*}
We consider $U = 3000$ users and $I = 1000$ items. We sample one item for each user over $30$ timesteps, where items are selected uniformly at random when $t=1$
and for $t > 1$ we have
\[
    P\left(A_{t, ui} = 1\right) \propto 
    \begin{cases}
        0& \text{ if user $u$ already interacted with item $i$}\\
        10& \text{ if } \mathrm{rank}_t(u, i) <= 100\\
        1& \text{ otherwise},
    \end{cases}
\]
where the ranking function $\mathrm{rank}_t(u, i)$ orders items from largest to smallest based on a score function intended to mimic the recommendation process: 
\[
    \mathrm{score}_t(u, i) \propto \sum_{s=1}^{t-1}\sum_{j=1}^I A_{s, uj} R_{s, uj} \exp{(S_{ij})},
\]
where $S_{ij}$ is an item-item similarity matrix with distribution
\[
    S_{ij} \sim \mathrm{Beta}'\left(\beta_i^{\top} \beta_j\right).
\]
We use the first $20$ sampled items for each user as the training set and do not consider the last $10$ for consistency with \citet{pan2021correcting}.\footnote{\citet{pan2021correcting} use half of the last $10$ items for evaluations and the other half as a validation set. This requirement does not apply to our algorithm: we do not need a validation set since we use the same hyperparameter settings as \citet{pan2021correcting}.} Finally, we sample an additional $20$ unobserved ratings uniformly at random for each user to create the test set.

We then train a generalized matrix factorization model \citep{he2017neural} using Adam with identical hyperparameter settings to \citet{pan2021correcting} but with each observation in the training loss weighted according to \gls{CAFL}. We then evaluate the recommender's predictions on the test set, repeating the entire simulation procedure $10$ times.
\begin{table}
\begin{center}
\begin{tabular}{ c|c|c } 
  Correction & MSE & MAE \\ 
 \hline
 Naive & $2.001 \pm 0.066$ & $1.087 \pm 0.021$ \\ 
 Pop& $1.990 \pm 0.035$ & $1.080 \pm 0.010$\\ 
 PF \citep{gopalan2015scalable}& $1.945 \pm 0.038$ & $1.065 \pm 0.010$ \\ 
 Pan \citep{pan2021correcting}& $1.896 \pm 0.042$ & $1.042 \pm 0.011$ \\ 
 \gls{CAFL} (This paper)& $\mathbf{1.818 \pm 0.019}$ & $\mathbf{1.015 \pm 0.005}$ \\ 
\end{tabular}
\caption{Predictive performance (MSE and MAE) of generalized matrix factorization on a benchmark derived from a modification of the simulation proposed by \citet{chaney2018algorithmic} when trained with: no correction (Naive), a correction that scales inversely with item popularity (Pop), Poisson Factorization (PF), the correction by \citet{pan2021correcting} (Pan), and the correction algorithm proposed in this work (\gls{CAFL}).}
\label{tab:comparison}
\end{center}
\end{table}

Table~\ref{tab:comparison} shows the performance of \gls{CAFL} averaged across all $10$ runs compared to the correction methods evaluated by \citet{pan2021correcting}. \gls{CAFL} outperforms all prior methods both in terms of MSE and MAE. We observe that the improvement in MSE/MAE when comparing \gls{CAFL} to Pan is larger than the improvement in MSE/MAE when comparing Pan to Poisson Factorization. Furthermore, we note that the MSE gap between the simple popularity-based re-weighting scheme and Pan is equal to the MSE gap between \gls{CAFL} and Pan, indicating that the \gls{CAFL} algorithm proposed in this work leads to significant performance improvements.

\section{Discussion}
\label{sec:discussion}

Feedback loops are endemic in multi-step recommender systems.  Recommendations affect user behavior; which in turn affect future recommendations through the retraining process. Feedback loops in recommender systems bias the inference of user preferences, compromise recommendation quality, and can homogenize recommendations. To this end, we propose \gls{CAFL}, a causal adjustment algorithm that can provably break feedback loops. Across empirical studies, we find that \gls{CAFL} improves recommendation quality and mitigates negative feedback effects. It also significantly improves predictive performance when compared to prior correction methods.

Furthermore, our results on homogenization show the importance of isolating feedback effects when evaluating models in dynamic setting. Our results indicate that the model's inductive bias and the number of datapoints, can sometimes have a stronger effect on homogenization than feedback loops. 
The picture of how and when homogenization occurs in recommender systems still remains incomplete. Future work that meticulously evaluates recommender systems in dynamic settings will likely shed light on this phenomenon.

\bibliographystyle{plainnat}
\bibliography{references}

\clearpage
\setcounter{page}{1}
\appendix
\onecolumn
{\Large\textbf{Appendix}}

\section{Proofs}

\subsection{Proof of \Cref{thm:ipw-positivity-violation}}
\label{app:causal-proof}

\begin{proof}
We decompose the causal objective into terms where positivity holds
and those where positivity is violated:
\begin{align}
&L^{\mathrm{causal}}(\Theta) \\
&= \frac{1}{t}\sum_{s=1}^t \sum_{u=1}^U\sum_{i=1}^I \E{\mbA_s}{\E{P(R_{s,ui}\g \mathrm{do}(A_{s,ui}=A_{s,ui}))}{\log P_\Theta\left(R_{s,ui}\g A_{s,ui}\right)}}\\
&=\sum_{s=1}^t c_t \left[\sum_{u=1}^U\sum_{i=1}^I \E{\mbA_s}{\E{P(R_{s,ui}\g \mathrm{do}(A_{s,ui}=A_{s,ui}))}{\log P_\Theta\left(R_{s,ui}\g A_{s,ui}\right)}}\right]\\
&=\sum_{s=1}^t c_t \left[\sum_{u=1}^U\sum_{i=1}^I\sum_{a_{s,ui}\in \{0,1\}} \E{P(R_{s,ui}\g \mathrm{do}(A_{s,ui}=a_{s,ui}))}{\mathbb{1}\{ A_{s,ui}=a_{s,ui} \}\cdot \log P_\Theta\left(R_{s,ui}\g A_{s,ui}\right)}\right]\\
&=\sum_{s=1}^t c_t \left[\sum_{\substack{u,i: \\P(A_{s,ui}=\\a_{s,ui}\g \hat{\Theta}_{s-1})>0}} \sum_{a_{s,ui}\in \{0,1\}} \E{P(R_{s,ui}\g \mathrm{do}(A_{s,ui}=a_{s,ui}))}{\mathbb{1}\{ A_{s,ui}=a_{s,ui} \}\cdot \log P_\Theta\left(R_{s,ui}\g A_{s,ui}\right)}\right] \nonumber \\
&\quad + \sum_{s=1}^t c_t \left[\sum_{\substack{u,i: \\P(A_{s,ui}=\\a_{s,ui}\g \hat{\Theta}_{s-1})=0}} \sum_{a_{s,ui}\in \{0,1\}} \E{P(R_{s,ui}\g \mathrm{do}(A_{s,ui}=a_{s,ui}))}{\mathbb{1}\{ A_{s,ui}=a_{s,ui} \}\cdot \log P_\Theta\left(R_{s,ui}\g A_{s,ui}\right)}\right].
\end{align}

The first equation is due to \Cref{eq:causal-objective}; the second
equation is due to the stationary assumption of intervention
distributions (i.e. the second assumption of
\Cref{thm:ipw-positivity-violation}); the third equation is an
unbiased estimator of the expectation over $\mbA_s$; the fourth
equation separates the loss into two terms, one where positivity holds
and the other where positivity fails.
$\hat{L}_{s,u,i}^\mathrm{obs}(\Theta\s a_{s,ui})$ in the theorem is an
unbiased estimator of the second term, following the same inverse
probability argument as in \Cref{prop:unbiased-causal,eq:ipw}.
$\hat{L}_{s,u,i}^\mathrm{unobs}(\Theta\s a_{s,ui})$ is an unbiased
estimator of the first term due to the stationary assumption of
intervention distributions, together with the inverse probability
argument as in \Cref{prop:unbiased-causal,eq:ipw}.
\end{proof}

\clearpage

\end{document}